\documentclass[11pt,peerreview,singlecolumn,double,draftclsnofoot,romanappendices]{IEEEtran}
\usepackage{pmat}
\usepackage{arydshln}
\usepackage[cmex10]{amsmath}
\usepackage{graphics}
\usepackage{mdwmath,mdwtab,url}
\usepackage[caption=false,font=footnotesize]{subfig}
\usepackage{stfloats}
\usepackage{nomencl}
\usepackage{amssymb}
\usepackage{leftidx}
\usepackage{rotating}
\usepackage{psfrag}
\usepackage{color}
\usepackage{hyperref}
\usepackage{cite}
\usepackage{epstopdf}
\usepackage{mathtools}

\newtheorem{theorem}{ {Theorem}}

\newtheorem{proposition}{{Proposition}}
\newtheorem{definition}{{Definition}}
\newtheorem{property}{\ti{Property}}
\newtheorem{lemma}{ {Lemma}}
\newtheorem{remark}{ {Remark}}

\newcommand{\mb}{\mathbb} 
\newcommand{\dv}{\mathbf} % vector in math bold font
\newcommand{\mc}{\mathcal} % variable in math calligraphic font
\newcommand{\ti}{\textit}
\newcommand{\tf}{\textbf}

\newcommand{\tc}{\textcolor}
\definecolor{brown}{rgb}{.75,.5,.5} 
\begin{document}
 
\title{Interplay Between Delayed CSIT and Network Topology for Secure MISO BC}

\author{Zohaib~Hassan~Awan and Aydin Sezgin 

\thanks{Zohaib Hassan Awan and Aydin Sezgin are with the Institute of  Digital Communication Systems, Ruhr-Universit\"{a}t Bochum, 44780 Bochum, Germany. Email: \{zohaib.awan, aydin.sezgin\}@rub.de}

%\thanks{This work has been supported by the European Commission in the framework of the FP7 Network of Excellence in Wireless Communications (NEWCOM\#), and the Concerted Research Action, SCOOP. The authors would also like to thank BELSPO for the support of the IAP BESTCOM  network.}

\thanks{ The results in this work was presented in part at the 10th International ITG Conference on Systems, Communications and Coding, Hamburg, Germany, Feb. 2015~\cite{scc}.}

%\thanks{\noindent \hspace{-.5em} This work is supported by the German Research Foundation, Deutsche
%Forschungsgemeinschaft (DFG), Germany, under grant SE 1697/11.}
}
\maketitle

\begin{abstract}
We study the problem of secure transmission over a Gaussian two-user multi-input single-output (MISO) broadcast channel under the assumption that links connecting the transmitter to the two receivers may have unequal strength \ti{statistically}. In addition to this, the state of the channel to each receiver is conveyed in a strictly causal manner to the transmitter. We focus on a two state topological setting of strong v.s. weak links. Under these assumptions, we first consider the MISO wiretap channel and establish bounds on generalized secure degrees of freedom (GSDoF). Next, we extend this model to the two-user MISO broadcast channel and establish inner and outer bounds on GSDoF region with different topology states.
The encoding scheme sheds light on the usage of both resources, i.e., topology of the model and strictly causal channel state information at the transmitter (CSIT); and, allows digitization and multi-casting of overheard side information, while transmitting confidential message over the stronger link. Furthermore, for a special class of channels, we show that the established bounds agree and so we characterize the sum GSDoF.   
\end{abstract}

\section{Introduction}\label{secI}
In communication networks, due to the scarcity of available resources and increase in the demand of higher data rates imposed by the consumers, multiple nodes communicate with each other over a shared medium. This in turn leads to a fundamental problem of interference in networks. A key ingredient to eradicate the detrimental effect of interference efficiently is by means of CSIT. In existing literature, for instance~\cite{J10}, different schemes are proposed which require perfect knowledge of CSIT to align or cancel interference. In practice, wireless medium is exposed to various random effects; thus, conveying perfect CSIT is difficult. Recently, in~\cite{M-AT12} Maddah-Ali \ti{et al.} study a  MISO broadcast channel and show a rather surprising result that strictly causal (delayed) CSIT is still useful in the sense that it enlarges the degrees of freedom (DoF) compared to a similar model with no CSIT. The model studied in~\cite{M-AT12} is generalized to a variety of settings namely, two- and three-user multi-input multi-output (MIMO) broadcast channel in~\cite{vaze_broadcast,abdoli}, two-user interference channel in~\cite{vaze_int,akbar_int}, and X-channel in~\cite{TMPS12a,akbar} all from DoF perspective. In cellular networks due to mobility, communication links are subjected to  different topological effects, e.g., inter-cell interference, wave propagation path loss, jamming. These physical factors influence links in an \ti{asymmetric} manner, that lead to some links being stronger than others \ti{statistically}.  A fundamental issue with  DoF analysis is that it ignores the diversity of links strength and implicitly assumes that all non-zero channels are equally strong in the sense that each link is capable of carrying 1 DoF, irrespective of the magnitude of channel coefficients. The GDoF metric solves this  limitation by taking diversity of links strength into account~\cite{JV10,Jafar14}. In~\cite{CEJ14}, Chen  \ti{et al.} study a two-user MISO broadcast channel by considering the two state topological setting of strong v.s. weak links and assume that CSI conveyed by both receivers can vary over time. For this model the authors establish bounds on GDoF region.
 
As said before, due to the broadcast nature of wireless medium, communication can be over heard by unintended nodes in the network. Wyner in~\cite{wyner}, introduced a basic wiretap channel to study secrecy by taking physical layer attributes of the channel into account. In Wyner's setup, the source wants to communicate a confidential message to the legitimate receiver and this message is meant to be concealed from the eavesdropper. For the degraded setting, in which the channel to legitimate receiver is stronger then to the eavesdropper secrecy capacity is established. In the last decade, the wiretap channel has attracted significant interest in the research community and is extended to study a variety of multi-user channels, e.g., the broadcast channel~\cite{csiszar,LLPS13}, the multi-access  channel~\cite{tekin,liangpoor,tekin2,Mac_2012,MAC_ieee}, the relay channel~\cite{lai,Z_allerton_2010,Z_relay_ieee}, the interference channel~\cite{onur_IFC,LYT08}, and the multi-antenna channel~\cite{khisti,oggier,Tie,BLPS09}. For a review of other related contributions the reader may refer to \cite{liangbook} (and references therein). Due to the difficulty in characterizing the complete secrecy capacity region, a number of recent contributions has focused on characterizing the \ti{approximate} capacity of these networks. The  approximate  capacity is measured by the notion of secure degrees of freedom (SDoF). Similar to the model with no security constraints, the SDoF metric captures the asymptotic behavior of secure data rates in high signal-to-noise ratio   (SNR) regime. Thus, SDoF can be equivalently understood as the secure spatial multiplexing gain, number of secure signaling dimensions, or the secrecy capacity pre-log factor.  From DoF perspective, the authors in~\cite{onur_dof} study a $K$-user interference channel and establish a lower bound on the sum SDoF, where perfect non-causal CSI is available at all nodes. Recently, Yang \textit{et al.} in~\cite{YKPS11} study the two-user MIMO broadcast channel under a relaxation that instead of non-causal CSI, strictly causal CSI (delayed) is provided to the transmitter from both receivers. For this model the authors characterize the SDoF region. The coding scheme in~\cite{YKPS11} follows by a careful extension of Maddah Ali-Tse scheme \cite{M-AT12} with additional noise injection to account for secrecy constraints. 
Zaidi \textit{et al.} in~\cite{Sdof-x-conference,Sdof-x} study the two-user MIMO X-channel with asymmetric feedback and delayed CSIT, and characterize the complete sum SDoF region. In \cite{Alternatin-isit}, the authors studied the MISO broadcast channel and assume that CSI conveyed by two receivers can vary over time and establish bounds on SDoF region. Recall that,  similar to DoF --- SDoF metric ignores the diversity of links strength at the receivers which may be beneficial to strengthen the secrecy in certain situations. Thus, going beyond the SDoF metric to the GSDoF will be useful to gain further insights and is the focus of this work.
\begin{figure}
\psfragscanon
\centering
\psfrag{x}{$\dv{x}$}
\includegraphics[width=.75\linewidth]{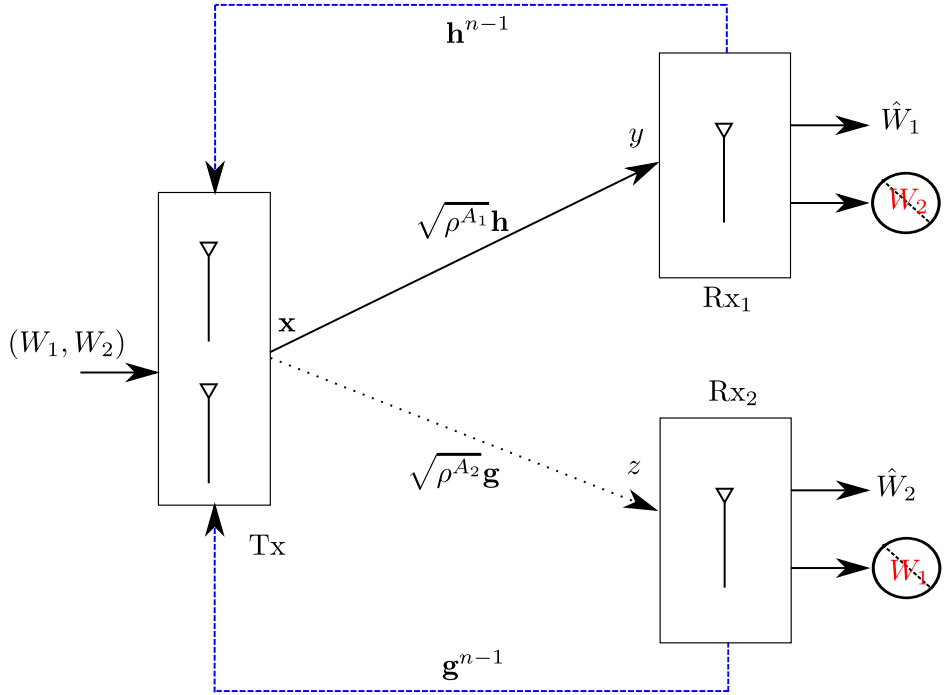}
\caption{$(2,1,1)$--MISO broadcast channel, where the link power exponent to receiver 1 is $A_1 \in \{1,\alpha\}$ and to receiver 2 is $A_2\in \{1,\alpha\}$ and $0 \le \alpha \le 1$.}
\psfragscanoff
\label{model}
\end{figure}
  
In this paper, we consider a Gaussian $(2,1,1)$--MISO broadcast channel which consists of three nodes --- a transmitter and two receivers as shown in Figure~\ref{model}. The transmitter is equipped with two antennas and each receiver is equipped with a single antenna. The transmitter wants to reliably transmit message $W_1$ to the receiver 1 and
message $W_2$ to the receiver 2. In investigating this model we make three assumptions, namely, 1) each receiver knows the perfect instantaneous CSI and also the CSI of the other receiver with a unit delay, 2) each receiver is allowed to convey the past or delayed CSI to the transmitter, and 3) links connecting two receivers may have different strength, statistically. We restrict our attention to the two state topological setting of stronger v.s. weaker links; thus, the topology of this network is allowed to alternate between four possible states and is known at the transmitter. Furthermore, message $W_1$ intended for the receiver 1 is meant to be kept secret from the receiver 2 and message $W_2$ intended for the receiver 2 is meant to be kept secret from the receiver 1. Thus, each receiver plays two different roles, not only 1) it acts as a legitimate receiver for the message intended for itself, 2) it also acts as an eavesdropper for the message intended for the other receiver. We assume that the eavesdroppers are passive and are not capable to modify the communication. 

The MISO broadcast channel that we study in this paper relates to a number of works studied previously. Compared to the MIMO broadcast channel with secrecy constraints studied in~\cite{YKPS11}, in this work the links connecting two receivers may observe different strength. The $(2,1,1)$--MISO broadcast channel that we study can be seen as a special case of the one in~\cite{CEJ14} but with imposed security constraints. From a practical viewpoint, the  channel shown in Figure~\ref{model} may be useful to model the down-link phase of a cellular network in which the base station wants to securely communicate with two receivers and the messages are meant to be kept secret from each other --- where both receivers are subjected to jamming from an external interferer.

The main contributions of this work are summarized as follows. We first consider the   $(2,1,1)$--MISO wiretap channel and establish bounds on the GSDoF. \iffalse The upper bound follows by extending the converse proof in~\cite{CEJ14} to account for secrecy constraints and that of~\cite{YKPS11} by taking topology of the network into account in a non-trivial manner.  We also establish lower bounds on allowed GSDoF with different topology states.\fi Next, we extend the MISO wiretap model to the broadcast setting with strictly causal (delayed) CSIT as shown in Figure~\ref{model}. For this model, we first establish an outer bound on the GSDoF region. The techniques used to establish the outer bound are essentially similar to the one that we use to prove the upper bound for the wiretap model. Concentrating on the role of topological diversity, we consider two elementary settings. In the first scenario, we consider a setting in which the link to one receiver is comparatively stronger than the other. We refer to this as \ti{fixed} topology. In the second scenario,  we consider the \ti{symmetric alternating} topology. This setting refers to the case in which the link to one receiver is stronger than to the other receiver, half of the duration of communication time. For these two models, we establish inner bounds on the GSDoF region. \iffalse, and, in doing so, we characterize the sum GSDoF.\fi The encoding scheme is based on an appropriate extension of Maddah Ali scheme~\cite{M-AT12} with noise injection~\cite{YKPS11}, and carefully utilizes the topology of the network. The key ingredients of the coding scheme are, as opposed to~\cite{M-AT12} where side information is conveyed in an analog manner, digitized side information is multicasted, and in supplement to this, fresh confidential information is send to the receiver with the stronger link. 

We also study a special class of $(2,1,1)$--MISO broadcast channel with integer channel coefficients. We first consider the MISO wiretap channel with fixed topology, where the link to legitimate receiver is stronger than to the eavesdropper and characterize the GSDoF. The coding scheme in this case follows by appropriately combining compute-and-forward scheme~\cite{nazar} and also uses some elements of the schemes that we have previously developed for the general case. Next, we consider the $(2,1,1)$--MISO broadcast channel with \ti{symmetric alternating} topology and characterize the sum GSDoF. Furthermore, we specialize our results for the case in which there are no security constraints. In particular, for the \ti{fixed} topology setting in which the link to one receiver is comparatively stronger than to the other, we characterize the  GDoF region. The coding scheme in this case follows by specializing the scheme that we have developed previously for a related model by removing the secrecy constraints. Finally, we illustrate our results with the help of some numerical examples.

We structure this paper as follows. Section~\ref{I} provides a formal description of the channel model along with some useful definitions. In Section~\ref{II}, we consider the MISO wiretap channel and state the upper and lower bounds on GSDoF. In Section~\ref{III}, we study the MISO broadcast channel and state the outer and inner bounds on the GSDoF region. In Section~\ref{special}, we specialize our results in previous sections to integer channels. Section~\ref{V} provides some numerical examples. Finally, in Section~\ref{conclusion} we conclude this paper by summarizing its contributions.

\vspace{.5em}

\textit{Notations:}
A few words about notations.  Boldface upper case letter $\dv X$ denote matrices, boldface lower case letter $\dv x$ denote vectors, and calligraphic letter $\mc X$ designate alphabets. At each time instant $t$, $\dv x_t$ denote $[x_{t1},\hdots,x_{tn}]$, and $\mathbb{E}[.]$ denote the expectation operator.  We use $\doteq$ to denote an exponential equality, such that given $f(\rho)\doteq \rho^{\beta}$  implies $\underset{\rho \rightarrow \infty}{\lim}\log f(\rho)/\log (\rho)=\beta$.  We use $\mc O (f(\rho))$ to denote the asymptotic behaviour of the function $f(\rho)$. The term $o(n)$ is some function $g(n)$ such that $\underset{n \rightarrow \infty}{\lim} \frac {g(n)}{n}=0$. The Gaussian distribution with mean $\mu$ and variance $\sigma^2$ is denoted by $\mc {CN}(\mu,\sigma^2)$. Finally, throughout the paper, logarithms are taken to base $2$.

\section{System Model and Definitions}\label{I}
We consider a two-user $(2,1,1)$--Gaussian MISO broadcast channel, as shown in Figure~\ref{model}. In this model, the transmitter is equipped with two transmit antennas and each of the receiver is equipped with a single antenna. The transmitter wants to reliably transmit message $W_{1} \in  \mc{W}_{1}=\{1,\hdots,2^{nR_{1}(A_1,\rho)} \}$ to receiver 1, and message $W_{2} \in \mc{W}_{2}=\{1,\hdots,2^{nR_{2}(A_2,\rho)} \}$ to receiver 2, respectively; and, in doing so, it wishes to conceal the message $W_{1}$, intended to receiver 1, from receiver 2  and the message $W_{2}$, intended to receiver 2, from receiver 1, respectively. Thus, receiver 2 not only is a legitimate receiver for confidential message $W_2$, it acts as an eavesdropper for the MISO channel to receiver 1. Similarly, receiver 1 not only is a legitimate receiver for confidential message $W_1$, it is an eavesdropper for the MISO channel to receiver 2.  For this setting, we consider a fast fading environment and assume that each receiver is fully aware of its own perfect instantaneous CSI and also the CSI of the other receiver with a unit delay. In addition to this, each receiver is allowed to convey only the past or outdated CSI to the transmitter, i.e., at time instant $t$, transmitter has perfect knowledge of \ti{only} the past $(t-1)$ channel states from both receivers. It is easy to see that by setting  $W_i=\phi$ for $i=1$ or $2$,  the model in Figure~\ref{model} reduces to the $(2,1,1)$--MISO wiretap channel.

Due to the inherent randomness of the wireless channel and topological changes that may arise, for instance --- due to the mobility of the users or interference (jamming) from unintended nodes, some elements of the network can experience more interference compared to the others. These factors in turn originate two fundamental classes of links, where few links are comparatively stronger than others  statistically. 
Let $A_1 \in \{1,\alpha\}$ denote the link power exponent from the  transmitter-to-receiver 1 and  $A_2 \in \{1,\alpha\}$ denote the link power exponent from transmitter-to-receiver 2, respectively, for $0 \le \alpha \le 1$; where we denote the stronger link by $A_i:=1$ and  weaker link by  $A_i:=\alpha$, $i=1,2$. As alluded before, the notion of stronger v.s. weaker links implies a statistical comparison, so for instance,  $A_1 > A_2$ refers to the case in which link connecting the transmitter to receiver 1 is stronger than to the receiver 2 statistically. For convenience, without loss of generality in the rest of the paper, we will refer to this as receiver 1 being \ti{stronger} than receiver 2. Then, based on the topology of the network, the model that we study belongs to any of the four possible states, $(A_1, A_2) \in \{1,\alpha\}^2$. We denote $\lambda_{A_1 A_2}$ be the fraction of time topology state $(A_1, A_2)$ occurs, such that
\begin{equation}
\label{sum}
\sum_{(A_1, A_2) \in \{1,\alpha\}^2}\lambda_{A_1 A_2} = 1.
\end{equation}
\noindent The channel input-output relationship at time instant $t$ is then given by
\begin{align}
\label{g-chan}
y_t &= \sqrt{\rho^{A_{1,t}}}{{\dv{h}}}_t {{\dv{x}}}_t+n_{1t} \notag\\
z_t &= \sqrt{{\rho}^{A_{2,t}}}{{\dv{g}}}_t {{\dv{x}}}_t+n_{2t}, \:\: t=1,\hdots,n
\vspace{-.5em}
\end{align}
where ${{\dv{x}}} \in \mb{C}^{2 \times 1}$ is the channel input vector, ${{\dv{h}}} \in \mc {H} \subseteq \mb{C}^{1 \times 2}$ is the channel vector connecting  receiver 1 to the transmitter and ${{\dv{g}}} \in \mc {G} \subseteq \mb{C}^{1 \times 2}$ is the channel vector connecting receiver 2 to the transmitter. The parameter $\rho$ is subject to input power constraint and the channel output noise $n_{i}$ is assumed to be independent and identically distributed (i.i.d.) white Gaussian noise, with $n_i \sim \mc{CN}(0,1)$ for $i= 1,2$. For convenience, we normalize the channel input vector, $||\dv{x}_t||^2 \leq 1$, then  the average received signal-to-noise ratio (SNR) for each link at time instant $t$  is given by 
\begin{eqnarray*}
 &&\mathbb{E}_{\dv{h}_t,\dv{x}_t}\big[||\sqrt{\rho^{A_{1,t}}}{{\dv{h}}}_t {{\dv{x}}}_t||^2\big] ={\rho^{A_{1,t}}}\\
&&  \mathbb{E}_{\dv{g}_t,\dv{x}_t}\big[||\sqrt{{\rho}^{A_{2,t}}}{{\dv{g}}}_t {{\dv{x}}}_t||^2\big] = {{\rho}^{A_{2,t}}}. 
\end{eqnarray*}
\noindent For ease of exposition, we denote $\dv{S}_t$ = $\left [
\begin{smallmatrix}
\dv{h}_t \\
\dv{g}_t
\end{smallmatrix} \right]$ as the channel state matrix and ${{\dv{S}}}^{t-1} =\{{{\dv{S}}}_1,\hdots,{{\dv{S}}}_{t-1}\}$ captures the collection of channel state matrices over the past $(t-1)$ symbols, respectively, where ${{\dv{S}}}^{0} =\emptyset$. We assume that, at each time instant $t$, the channel state matrix ${{\dv{S}}}_t$ is full rank almost surely. Furthermore, at each time instant $t$, the past states of the channel matrix $\dv{S}^{t-1}$ are known to all nodes. However, the instantaneous states $\dv{h}_t$ and $\dv{g}_t$ are known only to receiver 1, and receiver 2, respectively.

\vspace{.5em}
\begin{definition}\label{def1}
A code for the Gaussian two-user $(2,1,1)$--MISO broadcast channel with delayed CSIT and alternating topology consists of sequence of stochastic encoders at the transmitter,
\begin{align}
\label{map-PP}
\{\varphi_{t} \:\: &: \:\: \mc W_1 {\times} \mc W_2 {\times}\mc S^{t-1} \longrightarrow \mc{X}_{1} \times \mc{X}_2\}_{t=1}^{n}
\end{align}
where the messages $W_1$ and $W_2$ are drawn uniformly over the sets $\mc W_1$ and $\mc W_2$ respectively; and two decoding functions at receivers
\begin{align}
\psi_1 \:\: &: \:\: \mc Y^{n}{\times}\mc S^{n-1}{\times} \mc {H}_n \longrightarrow \hat{\mc{W}_1}\notag\\
\psi_2 \:\: &: \:\: \mc Z^{n}{\times}\mc S^{n-1}{\times} \mc {G}_n \longrightarrow \hat{\mc{W}_2}.
\end{align}
\end{definition}
 \vspace{.5em}
\begin{definition}\label{def2}
A rate pair $(R_1(A_1,\rho),R_2(A_2,\rho))$ is said to be achievable if there exists a sequence of codes such that
\begin{equation}
\limsup_{n \rightarrow \infty} \text{Pr}\{\hat{W}_{i} \neq W_{i} \}=0, \quad \forall\:\: i \in \{1,2\}.
\end{equation}
\end{definition}
 \vspace{.5em}
\begin{definition}\label{def3}
A GSDoF pair $(d_1(A_1),d_2(A_2))$ is said to be achievable if there exists a sequence of codes satisfying following
\begin{enumerate}
\item Reliability condition:
\begin{align}
& \limsup_{n \rightarrow \infty} \text{Pr}\{\hat{W}_{i} \neq W_{i}\}=0,\quad\quad \forall \:\: i \in \{1,2\},
\end{align}
\item Perfect secrecy condition:\footnote{For convenience, with a slight abuse in notations, we replace ${{\tf{S}}}^{n}:= (\tf{S}^{n-1}, \dv{h}_n)$,  ${{\tf{S}}}^{n}:= (\tf{S}^{n-1}, \dv{g}_n)$ in~\eqref{sec-constraint-1} and~\eqref{sec-constraint-2}, respectively.}
\begin{align}
\label{sec-constraint-1}
& \limsup_{n \rightarrow \infty} \frac{I(W_{2};y^n,{{\tf{S}}}^{n})}{n}=0, \\
\label{sec-constraint-2}
& \limsup_{n \rightarrow \infty} \frac{I(W_{1};z^n,{{\tf{S}}}^{n})}{n}=0,
\end{align}
\item and communication rate condition:
\begin{align}
&\lim_{\rho \rightarrow \infty} \liminf_{n \rightarrow \infty} \frac{\log |\mc W_i(n,\rho,A_i)|}{n\log \rho}\geq d_{i}(A_i), \quad \forall\:\: i \in \{1,2\}.
 \end{align}
\end{enumerate}
\end{definition}
 
\section{GSDoF of MISO wiretap channel with delayed CSIT}\label{II}
 \noindent In this section, we investigate the GSDoF of the MISO wiretap channel with delayed CSIT. Before proceeding to state the results, we first digress to provide a useful lemma which we will repetitively use in this work.
 \vspace{.5em}
\begin{lemma}
\label{lemma}
For the Gaussian MISO channel in~\eqref{g-chan}, following inequalities hold 
\begin{subequations}
\begin{align}
\label{l1}
h(y^n,z^n|\dv{S}^n) \dot{\le} 2h(z^n|\dv{S}^n)+n\lambda_{1\alpha}(1-\alpha) \log(\rho),\\
\label{l2}
h(y^n,z^n|\dv{S}^n)\dot{\le} 2h(y^n|\dv{S}^n)+n\lambda_{\alpha 1}(1-\alpha) \log(\rho),\\
\label{l3}
h(y^n|\dv{S}^n)\dot{\le} 2h(z^n|\dv{S}^n)+n\lambda_{1\alpha}(1-\alpha) \log(\rho),\\
\label{l4}
h(z^n|\dv{S}^n)\dot{\le} 2h(y^n|\dv{S}^n)+n\lambda_{\alpha 1}(1-\alpha) \log(\rho).
\end{align}
\end{subequations}
\end{lemma} 
 \vspace{.5em}
\begin{IEEEproof}
The proof of Lemma~\ref{lemma} appears in Appendix~\ref{app-1}. The inequalities in Lemma~\ref{lemma} also hold with additional conditioning over message $W$.  
\end{IEEEproof}
\iffalse
\vspace{.5em}
\begin{remark}
It is clear from~\eqref{g-chan} that, due to the \ti{asymmtery} of link power exponents at two receivers the entropy symmetry property of channel outputs~\cite{vaze_int,YKPS11},  may not hold in general. The proof follows by noticing that inequalities in~\cite[Lemma 1]{YKPS11}, however, do hold under \ti{local} output symmetry; and, extends it to account for imposed network topology.
\end{remark}
\fi
\subsection{Upper Bound}
 \noindent We now establish an upper bound on the GSDoF of the MISO wiretap channel with delayed CSIT and alternating topology.
  \vspace{.5em}
\begin{theorem}
\label{theorem-sdof-wt}
For the $(2,1,1)$--MISO wiretap channel with delayed CSIT and alternating topology $(\lambda_{A_1A_2})$, an upper bound on  GSDoF is given by
\begin{align}
d (\lambda_{A_1A_2}) & \le  \frac{(3-\alpha)\lambda_{1 \alpha}+2(\lambda_{11}+\alpha\lambda_{\alpha\alpha})+(1+\alpha)\lambda_{\alpha 1}}{3}.
\label{eq-sdof-wt}
\end{align}
\end{theorem}
 \vspace{.5em}
\begin{IEEEproof}
The proof of Theorem~\ref{theorem-sdof-wt} appears in  Appendix~\ref{proof-sdof-wt}.
\end{IEEEproof}
\iffalse
 \vspace{.5em}
\begin{remark}
The upper bound follows by extending the proof in~\cite{YKPS11} by taking network topology into account. By removing the topology consideration, i.e., setting $\lambda_{11}:=1$ in~\eqref{eq-sdof-wt}, the upper bound reduces to the SDoF of the MISO wiretap channel with delayed CSI \cite[Theorem 1]{YKPS11}.
\end{remark}
\fi

\subsection{Coding schemes with fixed topology}
Next, we provide some encoding schemes for fixed topology states. For simplicity of analysis and in accordance with DoF framework, in this work we neglect the additive Gaussian noise and only mention the asymptotic behavior of the inputs by ignoring the exact power allocations.

\vspace{.5em}
\subsubsection{Fixed Topology ($\lambda_{1\alpha}=1$)}
\label{fixed-topology}
We now focus our attention on the case in which receiver 1 (legitimate receiver) is stronger than receiver 2 (eavesdropper), comparatively and state a lower bound on  GSDoF. From practical viewpoint, this case may be useful to model a setting in which the legitimate receiver is geographically located at a more favorable position compared to the eavesdropper, and observes less interference from an external interferer (jammer) as opposed to the eavesdropper.

\vspace{.5em}
\begin{proposition}
\label{prop1}
The GSDoF  of  $(2,1,1)$--MISO wiretap channel
with delayed CSIT and fixed topology ($\lambda_{1\alpha}=1$)  is given by   
\begin{eqnarray}
\label{prop1-eq}
\frac{2}{3} \le d \le 1- \frac{\alpha}{3}.
\end{eqnarray}
\end{proposition}
 \vspace{.5em}
\begin{proof}
\begin{figure}
  \psfragscanon
  \centering
  \psfrag{a}{$\rho$}
  \psfrag{j}{$\rho^{1-\alpha}$}
  \psfrag{i}{\hspace{-.5em}$\rho^{\alpha}$}
  \psfrag{c}{$\tc{green}{v_1}$}
  \psfrag{b}{\hspace{-.5em}$\tc{red}{\dv{h}_1\dv{u}}$}
  \psfrag{d}{\hspace{-.25em}$\tc{red}{\dv{g}_1\dv{u}}$}
  \psfrag{e}{\hspace{-.25em}$\dv{h}_2\tc{green}{\dv{v}}$}
  \psfrag{+}{\hspace{.25em}$+$}
  \psfrag{f}{\hspace{-1.1em}$h_{21}\tc{red}{\dv{h}_1\dv{u}}$}
  \psfrag{g}{\hspace{-.25em}$\dv{g}_2\tc{green}{\dv{v}}$}
  \psfrag{h}{\hspace{- 1.1em}$g_{21}\tc{red}{\dv{h}_1\dv{u}}$}
  \psfrag{k}{\hspace{- 1.4em}$h_{31}(\dv{g}_2\tc{green}{\dv{v}}$}
  \psfrag{l}{\hspace{- 1.4em}$g_{21}\tc{red}{\dv{h}_1\dv{u}})$}
  \psfrag{m}{\hspace{- 1.2em}$g_{31}(\dv{g}_2\tc{green}{\dv{v}}$}
  \psfrag{n}{\hspace{- 1.1em}$g_{21}\tc{red}{\dv{h}_1\dv{u}})$}
  \psfrag{p}{ $\tc{green}{v_1}$}
  \psfrag{q}{$\rho^{0}$}
  \psfrag{t1}{$t=1$}
  \psfrag{t2}{$t=2$}
  \psfrag{t3}{$t=3$}
  \psfrag{R1}{\hspace{-5em}Received power levels at Rx$_1$}
  \psfrag{R2}{\hspace{-5em}Received power levels at Rx$_2$}
  \includegraphics[scale=1]{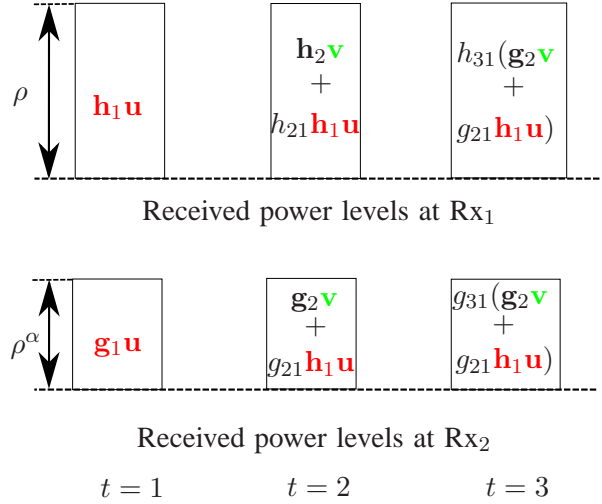}
  \caption{Received power levels at receiver 1 and receiver 2.}
  \psfragscanoff
  \label{power-level-base}
  \end{figure}
The upper bound follows immediately from the bound established in Theorem~\ref{theorem-sdof-wt} by setting $\lambda_{1\alpha}:=1$ in~\eqref{eq-sdof-wt}. We now provide the description of the encoding scheme that we use to prove the lower bound in Proposition~\ref{prop1}.  The coding scheme in this case is an adaptation of the scheme developed by~\cite{YKPS11}, and, so we outline it  briefly.  In this scheme, the transmitter sends two symbols $(v_1,v_2)$ to receiver 1 and wishes to conceal them from receiver 2. The coding scheme consists of three time slots. In the first time slot the transmitter injects uncoded Gaussian noise $(\dv{u}:=[u_1, u_2]^T)$ from both antennas. Note that, due to the topology of the network, the channel output at receiver 2 (eavesdropper) is available at a lower power level compared to the  receiver 1  as shown  in Figure~\ref{power-level-base}. The channel output at the receiver 1 can be interpreted as a secret key that helps to secure the confidential messages in the next timeslot. At the end of timeslot 1, by means of past CSI the transmitter can \ti{learn} the channel output at receiver 1 and sends it with confidential symbols $(\dv{v}:=[v_1, v_2]^T)$ intended for receiver 1. At the end of second timeslot, receiver 1 gets the confidential symbols embedded in with secret key $(\dv{h}_1\dv{u})$. Since it knows the CSI and $\dv{h}_1\dv{u}$, it subtracts out the contribution of $\dv{h}_1\dv{u}$ from the channel output at the end of second timeslot to get one equation with two variables and  requires one extra equation to decode the confidential symbols, which is being available as side information at receiver 2. By means of past CSI, the transmitter can construct the channel output at receiver 2 in timeslot 2 and in the third time slot sends it to the receiver 1, that helps to decode the two symbols securely.
\end{proof}
 \vspace{.5em}
\begin{remark}
 It can be easily seen  from~\eqref{prop1-eq} that the lower and upper bounds do not coincide in general. However, for a special class of channels where the channel coefficients belong to $\dv {h}, \dv{g} \in \mb {Z}^{1 \times 2}$, we show that by combining some elements of this scheme with compute-and-forward scheme~\cite{nazar} the bounds agree and GSDoF is characterized. The special case will be discussed in Section~\ref{special}.
\end{remark}

\vspace{.5em}
\subsubsection{Fixed Topology ($\lambda_{\alpha 1}=1$)}
\label{fixed-topology-3}
This setting refers to the case in which link connecting transmitter-to-legitimate receiver is statistically weaker than to the eavesdropper, for which $2\alpha/3$  SDoF is achievable. The coding scheme follows along similar lines as in Proposition~\ref{prop1} and is omitted.
\vspace{.5em}
\iffalse
\subsubsection{No Topology ($\lambda_{11}=1$)}
\label{fixed-topology-1}
 By removing the topology consideration, i.e., by setting $(A_{1t},A_{2t}):=(1,1)$ $\forall \: t$, the model reduces to the MISO wiretap channel with delayed CSIT studied in~\cite{YKPS11}, for which  the optimal SDoF is given by $2/3$ SDoF. Alternatively, one can also recover the $2/3$ SDoF from the encoding scheme developed in Proposition~\ref{prop1}  by setting $\alpha:=1$ in~\eqref{prop1-eq}.
 \fi
\section{GSDoF of MISO broadcast channel with delayed CSIT}
\label{III}
Next, we extend the MISO wiretap channel model to the two-user broadcast setting and establish bounds on GSDoF region. 
\vspace{.5em}
\subsection{Outer Bound}

\noindent The following theorem provides an outer bound on the GSDoF region of the MISO broadcast channel with delayed CSIT.
\vspace{.5em}
\begin{theorem}
\label{theorem-sdof-bc}
For the $(2,1,1)$--MISO broadcast channel with delayed CSIT and alternating topology $(\lambda_{A_1A_2})$, an outer bound on  GSDoF region $\mc C_\text{GSDoF}(\lambda_{A_1 A_2})$ is given by the set of all non-negative pairs $(d_1,d_2)$ satisfying
\setlength{\arraycolsep}{0.2em}
\begin{subequations}
\label{outer-bound}
\begin{eqnarray}
\label{m1-bc}
 3d_1+d_2 &\le& (3-\alpha)\lambda_{1 \alpha}+2(\lambda_{11}+\alpha\lambda_{\alpha\alpha})+(1+\alpha)\lambda_{\alpha 1}\\
\label{m2-bc}
  d_1+3 d_2&\le& (3-\alpha)\lambda_{\alpha 1}+2(\lambda_{11}+\alpha\lambda_{\alpha\alpha})+(1+\alpha)\lambda_{1 \alpha }.
\end{eqnarray}
\end{subequations}
\setlength{\arraycolsep}{0.5em}
%double Coulmn
\iffalse
\begin{subequations}
\begin{eqnarray}
\label{m1-bc}
&& \hspace{-4em} 3d_1+d_2 \le (3-\alpha)\lambda_{1 \alpha}+2(\lambda_{11}+\alpha\lambda_{\alpha\alpha})+(1+\alpha)\lambda_{\alpha 1}\\
\label{m2-bc}
 &&\hspace{-4em} d_1+3 d_2\le (3-\alpha)\lambda_{\alpha 1}+2(\lambda_{11}+\alpha\lambda_{\alpha\alpha})+(1+\alpha)\lambda_{1 \alpha }.
\end{eqnarray}
\end{subequations}
\fi
\end{theorem}
\vspace{.5em}
\begin{IEEEproof}
The proof of Theorem~\ref{theorem-sdof-bc} appears in Appendix~\ref{proof-sdof-bc}. 
\end{IEEEproof}
 \vspace{.5em}
\subsection{Coding schemes with fixed topology}
\label{special-base}
\begin{table}
\centering 
\renewcommand{\arraystretch}{.75}
\begin{tabular}{| c | c | c | c | c |}
\hline 
Timeslot & $1$ & $2$ & $3$ & $4$\\ \hline 
 $\dv{x}$ & $\dv{u}$ & $ [v_1\: v_2]^T + L_1 (\dv{u})$ & $[w_1\: w_2]^T + M_1 (\dv{u})$ & $\begin{aligned} L_3( w_1,w_2, M_1 (\dv{u}))\\  +M_2( v_1,v_2, L_1 (\dv{u}))\end{aligned}$\\ \hline
Rx$_1$ & $y_1=\sqrt{\rho} L_1 (\dv{u})$ & $y_2=\sqrt{\rho} L_2( v_1,v_2, L_1 (\dv{u}))$ & $y_3=\sqrt{\rho} L_3( w_1,w_2, M_1 (\dv{u}))$  & $\begin{aligned}y_4=\sqrt{\rho}L_3( w_1,w_2, M_1 (\dv{u}))\\+\sqrt{\rho}M_2( v_1,v_2, L_1 (\dv{u}))\end{aligned}$ \\ \hline
Rx$_2$ & $z_1= \sqrt{\rho^{\alpha}} M_1 (\dv{u})$ & $z_2=\sqrt{\rho^{\alpha}} M_2( v_1,v_2, L_1 (\dv{u}))$ & $z_3=\sqrt{\rho^{\alpha}}M_3( w_1,w_2, M_1 (\dv{u}))$ & $\begin{aligned}z_4=\sqrt{\rho^{\alpha}}L_3( w_1,w_2, M_1 (\dv{u})\\+\sqrt{\rho^{\alpha}}M_2( v_1,v_2, L_1 (\dv{u}))\end{aligned}$\\ \hline
\end{tabular}
\vspace{1em}
\caption{Yang~\ti{et al.} scheme for fixed topology~$(\lambda_{1 \alpha}=1)$.}
\label{table}
\end{table} 
\noindent We now consider the $(2,1,1)$--MISO broadcast channel with fixed topology  $(\lambda_{1 \alpha}=1$ or $\lambda_{\alpha 1}= 1)$, where one receiver is comparatively  stronger than other. For sake of completeness, before providing the inner bound, we first revisit the scheme developed by {Yang~\ti{et al.}}~\cite[section V-B]{YKPS11} for the model without topology consideration $(A_{1 t},A_{2 t})=(1,1)$ to the fixed topology setting, i.e.,  $(A_{1 t},A_{2 t})=(1,\alpha)$,  $\forall$ $t$. A trivial inner bound on the GSDoF region $\mc C_\text{GSDoF}(\lambda_{A_1 A_2})$ of the two-user $(2,1,1)$--MISO broadcast channel with fixed topology  $(\lambda_{1 \alpha}=1)$ is given by the set of all non-negative pairs $(d_1,d_2)$ satisfying
\begin{subequations}
\label{trivial-sdof-miso-bc-outer-bound-eqautions}
\begin{align}
\label{t-miso-bc-outer-a}
3\alpha d_1+d_2 &\le 2\alpha \\
\label{t-miso-bc-outer-b}
   \alpha d_1+3d_2 &\le 2\alpha. 
\end{align}
\end{subequations}
The achievable GSDoF region is  given by the corner points  $(0,2\alpha/3)$, $(2/3,0)$, and the point $(1/2,\alpha/2)$ obtained by the intersection of line equations in~\eqref{trivial-sdof-miso-bc-outer-bound-eqautions}. The corner points $(2/3,0)$ and $(0,2\alpha/3)$ are readily achievable by using the coding scheme as in Sections~\ref{fixed-topology} and~\ref{fixed-topology-3}, since the model reduces to the equivalent $(2,1,1)$--MISO wiretap channel --- where the transmitter wants to send confidential message to one receiver and wishes to conceal it from the unintended receiver. We now give a sketch of the transmission scheme that is used to achieve the point $(1/2,\alpha/2)$. In this scheme, transmitter wants to transmit two symbols $(\dv{v}:=[v_1, v_2]^T)$ to receiver 1 and wishes to conceal them from receiver 2; and two symbols $(\dv{w} :=[w_1,w_2]^T)$ to receiver 2 and wishes to conceal them from receiver 1. The communication takes place in four phases, each comprising of only one timeslot. The encoding scheme is concisely shown in Table~\ref{table}. In the first timeslot the transmitter injects artificial noise $(\dv{u}:=[u_1, u_2]^T)$, where each receiver gets a linear combination of artificial noise denoted by $L_1(\dv{u})$ and $M_1(\dv{u})$ but with different power levels, respectively. Due to the availability of strictly causal or delayed CSI, at the end of time slot 1 the transmitter can learn the channel output at both receivers and in the second timeslot transmits fresh information $(\dv{v}:=[v_1, v_2]^T)$ to receiver 1 along with the channel output at receiver 1 in timeslot 1, $y_1$. At the end of timeslot 2, receiver 1 requires one equation to decode the intended symbols, being available as side information at receiver 2 $(z_2)$.  In the third timeslot, transmission scheme is similar to the one in timeslot 2 with the roles of receiver 1 and receiver 2 being reversed. In this phase, the transmitter sends fresh information $(\dv{w}:=[w_1, w_2]^T)$ to receiver 2 along with the past channel output at receiver 2 in timeslot 1 ($z_1$). By means of past CSIT, at the end of third time slot the transmitter can learn the side information at receiver 1 in timeslot 3 $(y_3)$ and at receiver 2 in timeslot 2 $(z_2)$, and sends them in timeslot 4. At the end of fourth time slot, since  receiver 1 knows the CSI, it first subtracts out the contribution of side information $y_3$ seen at receiver 1 from channel output $y_4$ to get $z_2$. Afterwards, it  removes the contribution of $y_1$ from $(y_2,z_2)$ and decodes $\dv v$ through channel inversion. Receiver 2 can also perform similar operations to decode $\dv w$.

\begin{figure}
\psfragscanon
\centering
\psfrag{a}[c][c]{\hspace{-.10em}$\frac{2}{3}$}
\psfrag{b}[c][c]{\hspace{-.35em} $1-\frac{\alpha}{3}$}
\psfrag{x}[c][c]{ $d_1(\alpha)$}
\psfrag{y}[c][c]{\hspace{-1em} $d_2(\alpha)$}
\psfrag{c}[c][c]{$\frac{2\alpha}{3} $}
\psfrag{d}{$\frac{1+\alpha}{3}$}
\psfrag{f}{$(1-\frac{\alpha}{2},\frac{\alpha}{2})$}
\psfrag{e}{$( \frac{1}{2},\frac{\alpha}{2})$}
\psfrag{k}{$\big(\frac{2}{3+\alpha},\frac{\alpha(1+\alpha)}{3+\alpha}\big)$}
\psfrag{g}{Outer bound~\eqref{outer-bound}}
\psfrag{i}{Yang \ti{et al.} Scheme~\eqref{trivial-sdof-miso-bc-outer-bound-eqautions}}
\psfrag{h}{Inner bound~\eqref{prop-2-eqautions}}
\includegraphics[width=.6\linewidth]{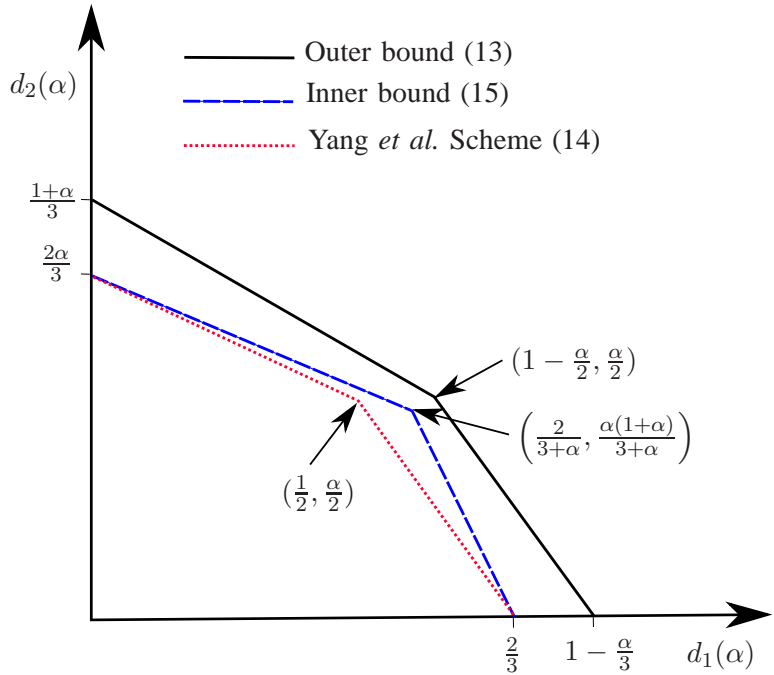}
\caption{Bounds on the GSDoF region of $(2,1,1)$--MISO broadcast channel, where the link to receiver 2 is weaker than to receiver 1 $(\lambda_{1 \alpha}=1)$.}
\psfragscanoff
\label{R-comp-sec}
\end{figure}

The equivocation analysis of this scheme, by proceeding
as in Appendix~\ref{equiv}, reveals that $2$ symbols are securely transmitted over a total of four time slots yielding GSDoF of $2/4$ at receiver 1. Due to the symmetry of the problem, it can be readily shown that $2$ symbols are securely send to receiver 2; where, due to the topology of the communication model each symbol is capable of carrying only $\alpha$ bits, which yields  GSDoF of $2\alpha/4$  at receiver 2. In Figure~\ref{R-comp-sec}, we plot the outer~\eqref{outer-bound} and inner bounds~\eqref{trivial-sdof-miso-bc-outer-bound-eqautions} for the two-user MISO broadcast channel with fixed topology. It can be easily seen from Figure~\ref{R-comp-sec} that the bounds in~\eqref{outer-bound} and~\eqref{trivial-sdof-miso-bc-outer-bound-eqautions} do not agree in general. In what follows, we provide an alternative coding scheme which gives an improved inner bound on GSDoF region. \iffalse and is shown to be optimal in certain cases.\fi
 
The following proposition gives an inner bound on the GSDoF region of the MISO broadcast channel with fixed topology  $(\lambda_{1 \alpha}=1)$.
\vspace{.5em} 
\begin{proposition}
\label{prop2}
An inner bound on the GSDoF region $\mc C_\text{GSDoF}(\lambda_{A_1 A_2})$ of the two-user $(2,1,1)$--MISO broadcast channel with delayed CSIT and fixed topology  $(\lambda_{1 \alpha}=1)$ is given by the set of all non-negative pairs $(d_1,d_2)$ satisfying
\begin{subequations}
\label{prop-2-eqautions}
\begin{align}
\label{miso-bc-outer-a}
3(1+\alpha)d_1+2d_2 &\le 2(1+\alpha) \\
\label{miso-bc-outer-b}
 \alpha(3-\alpha)d_1+6d_2 &\le 4\alpha. 
\end{align}
\end{subequations}
\end{proposition}
\begin{IEEEproof}
The inner bound follows by generalizing the coding scheme developed in Proposition~\ref{prop1} for the wiretap channel to the broadcast setting. \iffalse and also borrows some elements from the one in~\cite{CEJ14} by accounting for secrecy constraints.\fi The region in~\eqref{prop-2-eqautions} is characterized by the corner points $(2/3,0)$, $(0,2\alpha/3)$ and the point $(2 / (3+\alpha),\alpha(1+\alpha)/(3+\alpha))$ obtained by the intersection of line equations in~\eqref{prop-2-eqautions}. The achievability of the two corner points $(2/3,0)$  and  $(0,2\alpha/3)$ follow by the coding scheme developed in 
Proposition~\ref{prop1} and in section~\ref{fixed-topology-3}, respectively, where the transmitter is interested to send  message to only one receiver and the other receiver acts as an eavesdropper. The achievability of the point  $(2 / (3+\alpha),\alpha(1+\alpha)/(3+\alpha))$  is provided in Appendix~\ref{app-4}.
\end{IEEEproof}
 \vspace{.5em}
\begin{remark}
As seen in Figure~\ref{R-comp-sec}, the outer~\eqref{outer-bound} and  inner~\eqref{prop-2-eqautions} bounds do not meet in general; however, it is worth noting that the encoding scheme we have established in Proposition~\ref{prop2} provides a larger sum GSDoF compared to Yang \ti{et al.} scheme~\eqref{trivial-sdof-miso-bc-outer-bound-eqautions}, i.e., 
\begin{eqnarray}
 \underbrace{\frac{1+\alpha}{2}}_{\text {sum GSDoF}~\eqref{trivial-sdof-miso-bc-outer-bound-eqautions}}  \le \underbrace{\frac{2+\alpha(1+\alpha)}{3+\alpha}}_{\text {sum SDoF}\eqref{prop-2-eqautions}} 
\end{eqnarray}

\end{remark}

\subsection{Coding scheme with symmetric alternating topology}
 \begin{figure}
\psfragscanon
\centering
\psfrag{b}[c][c]{\hspace{-.5em} $\frac{2}{3}$}
\psfrag{a}[c][c]{  $\frac{1+\alpha}{3}$}
\psfrag{x}[c][c]{ $d_1(\alpha)$}
\psfrag{y}[c][c]{\hspace{-1em} $d_2(\alpha)$}
\psfrag{c}[c][c]{$\frac{1+\alpha}{3}$}
\psfrag{d}{\hspace{.5em}$\frac{2}{3}$}
\psfrag{f}{$( \frac{1}{2},\frac{1}{2})$}
\psfrag{e}{$( \frac{1+\alpha}{4},\frac{1}{2})$}
\psfrag{k}{$\frac{1}{2}$}
\psfrag{g}{Outer bound~\eqref{outer-bound}}
\psfrag{h}{Inner bound~\eqref{theorem-sdof-sym}} 
\includegraphics[width=.6\linewidth]{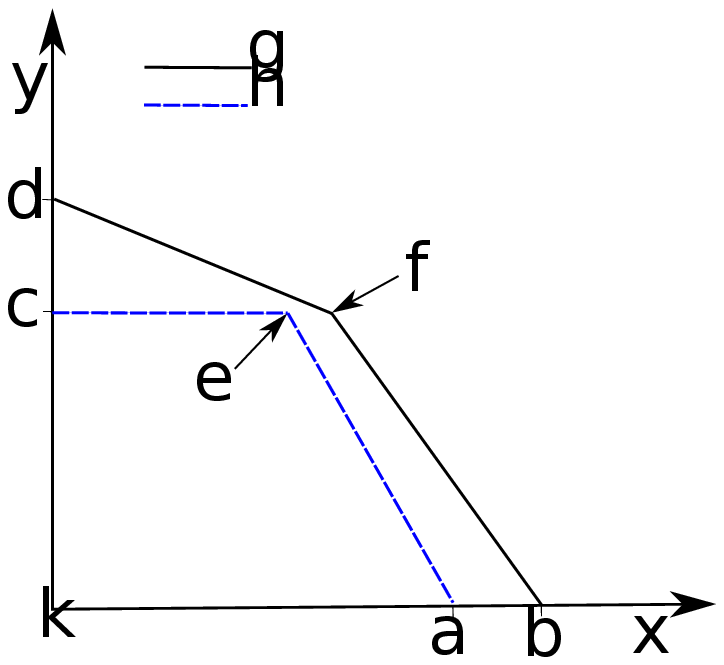}
\caption{Bounds on the GSDoF region of $(2,1,1)$--MISO broadcast channel with symmetric  alternating topology $(\lambda_{1 \alpha},\lambda_{\alpha 1})=( \frac{1}{2},\frac{1}{2})$.}
\psfragscanoff
\label{R-Comp-sym-sec-psfrag-base}
\end{figure} 

\noindent We now turn our attention to the MISO broadcast channel with alternating topology and focus on the symmetric case where the fractions of time spent in state $\lambda_{1\alpha}$ and $\lambda_{\alpha1}$ are equal. This communication channel may model a setting in which due to the mobility of the users each receiver experiences strong interference from an external jammer half of the duration of communication time. The following proposition provides an inner bound on GSDoF region of the MISO broadcast channel with symmetric alternating topology  $(\lambda_{1 \alpha},\lambda_{\alpha 1})= (\frac{1}{2}, \frac{1}{2})$.
\vspace{.5em} 
\begin{proposition}
\label{prop4}
An inner bound on the  GSDoF region $\mc C_\text{GSDoF}(\lambda_{A_1 A_2})$ of the two user $(2,1,1)$--MISO broadcast channel with delayed CSIT and symmetric alternating topology  $(\lambda_{1 \alpha}=\lambda_{\alpha 1}= \frac{1}{2})$ is given by the set of all non-negative pairs $(d_1,d_2)$ satisfying
\begin{subequations}
\label{theorem-sdof-sym}
\begin{align}
\label{miso-bc-outer-a}
6d_1+(1+\alpha) d_2 &\le 2(1+ \alpha)\\
\label{miso-bc-outer-b}
 2(2\alpha-1) d_1+3(1+\alpha)d_2 &\le (1+ \alpha)^{2}. 
\end{align}
\end{subequations}
\end{proposition}
\vspace{.5em} 
\begin{IEEEproof}
The inner bound follows by specializing the coding scheme that we establish for the fixed topology setting to the alternating topology setting. As seen from Figure~\ref{R-Comp-sym-sec-psfrag-base}, it is sufficient to prove the SDoF pairs $(\frac{1+\alpha}{3},0)$,  $(0,\frac{1+\alpha}{3})$ and $(\frac{1+\alpha}{4},\frac{1}{2})$, since the entire region~\eqref{theorem-sdof-sym} can then be achieved by time sharing. The SDoF pairs $(\frac{1+\alpha}{3},0)$ and $(0,\frac{1+\alpha}{3})$ are readily achievable by using the combination of coding schemes developed in Proposition~\ref{prop1} and in Section~\ref{fixed-topology-3}, equal fractions of time. The achievability of the  point $(\frac{1+\alpha}{4},\frac{1}{2})$ is relegated to Appendix~\ref{app-6}.
\end{IEEEproof}
\vspace{.5em}
\begin{remark}[Generalized sum SDoF Gains with Topological Diversity]
Proposition~\ref{prop4} provides an inner bound on the GSDoF region of MISO broadcast channel with symmetric alternating topology, where each receiver observes a strong link half of the duration of communication time. We note that sum GSDoF in Proposition~\ref{prop4} can be larger than the one obtained by a similar model but with \ti{non-diverse} topology consideration. For non-diverse topology setting, i.e., $\lambda_{11}$, $\lambda_{\alpha \alpha}$, the optimal sum GSDoF is given by 1 and $\alpha$, respectively~\cite{YKPS11}. The non-diverse topology model of $(\lambda_{11},\lambda_{\alpha \alpha})=(\frac{1}{2},\frac{1}{2})$ is equivalent to the set-up that we consider in Proposition~\ref{prop4} in the sense that the duration of communication time for stronger and weaker links for both receivers are same. The sum GSDoF with non-diverse topology is given by 
\begin{eqnarray}
\text{GSDoF} &=& \frac{1}{2} \times \underbrace{1}_{\text {sum GSDoF}\:  {(A_1,A_2)=(1,1)}} + \frac{1}{2} \times \underbrace{\alpha}_{\text {sum GSDoF}\:  {(A_1,A_2)=(\alpha,\alpha)}} \notag\\
&=& \frac{1+\alpha}{2} \le \underbrace{\frac{3+\alpha}{4}}_{\text {sum GSDoF}\:  (\lambda_{1 \alpha},\lambda_{\alpha 1})= (\frac{1}{2}, \frac{1}{2})} 
\end{eqnarray}
which is clearly smaller than the sum GSDoF of Proposition~\ref{prop4}. This result shows the benefits of topological diversity.
\end{remark}

\section{GSDoF (GDoF) Characterization in Few Special Cases}
\label{special}
 In this section, we study the two user Gaussian MISO broadcast channel~\eqref{g-chan} where we restrict our attention to integer channels, i.e., $\dv{h} \in \mc{H} \subseteq \mb Z^{1 \times 2 }$ and $\dv{g}\in \mc{G} \subseteq \in \mb Z^{1 \times 2 }$. \footnote{The results established for the integer channels can be readily extended to hold for complex channels using standard techniques~\cite{nazar,niesen}.} In what follows, we construct some elemental encoding schemes which characterize the (sum) GSDoF for various topology states. 
 \vspace{.5em}
 \subsection{Wiretap channel with Fixed Topology ($\lambda_{1\alpha}=1$)}
 \label{fixed-topology-new}
 We first consider the MISO wiretap channel with fixed topology state $(\lambda_{1\alpha}=1)$ and establish the optimal GSDoF. 
 \vspace{.5em}
 \begin{proposition}
 \label{prop1-new}
 The GSDoF  of  $(2,1,1)$--MISO wiretap channel
 with delayed CSIT and fixed topology state ($\lambda_{1\alpha}=1$) is given by   
 \begin{eqnarray}
 \label{prop1-eq-new}
 d = 1- \frac{\alpha}{3}.
 \end{eqnarray}
 \end{proposition}
  \vspace{.5em}
 \begin{IEEEproof} 
   \begin{figure}
   \psfragscanon
   \centering
   \psfrag{a}{$\rho$}
   \psfrag{j}{$\rho^{1-\alpha}$}
   \psfrag{i}{\hspace{-.5em}$\rho^{\alpha}$}
   \psfrag{c}{$\tc{green}{v_1}$}
   \psfrag{b}{\hspace{-.5em}$\tc{red}{\dv{h}_1\dv{u}}$}
   \psfrag{d}{\hspace{-.25em}$\tc{red}{\dv{g}_1\dv{u}}$}
   \psfrag{e}{\hspace{-.25em}$\dv{h}_2\tc{green}{\dv{v}}$}
   \psfrag{+}{\hspace{.25em}$+$}
   \psfrag{f}{\hspace{-1.1em}$h_{21}\tc{red}{\dv{h}_1\dv{u}}$}
   \psfrag{g}{\hspace{-.25em}$\dv{g}_2\tc{green}{\dv{v}}$}
   \psfrag{h}{\hspace{- 1.1em}$g_{21}\tc{red}{\dv{h}_1\dv{u}}$}
   \psfrag{k}{\hspace{- 1.4em}$h_{31}(\dv{g}_2\tc{green}{\dv{v}}$}
   \psfrag{l}{\hspace{- 1.4em}$g_{21}\tc{red}{\dv{h}_1\dv{u}})$}
   \psfrag{m}{\hspace{- 1.2em}$g_{31}(\dv{g}_2\tc{green}{\dv{v}}$}
   \psfrag{n}{\hspace{- 1.1em}$g_{21}\tc{red}{\dv{h}_1\dv{u}})$}
   \psfrag{p}{ $\tc{green}{v_1}$}
   \psfrag{q}{$\rho^{0}$}
   \psfrag{t1}{$t=1$}
   \psfrag{t2}{$t=2$}
   \psfrag{t3}{$t=3$}
   \psfrag{R1}{\hspace{-5em}Received power levels at Rx$_1$}
   \psfrag{R2}{\hspace{-5em}Received power levels at Rx$_2$}
   \includegraphics[scale=1.15]{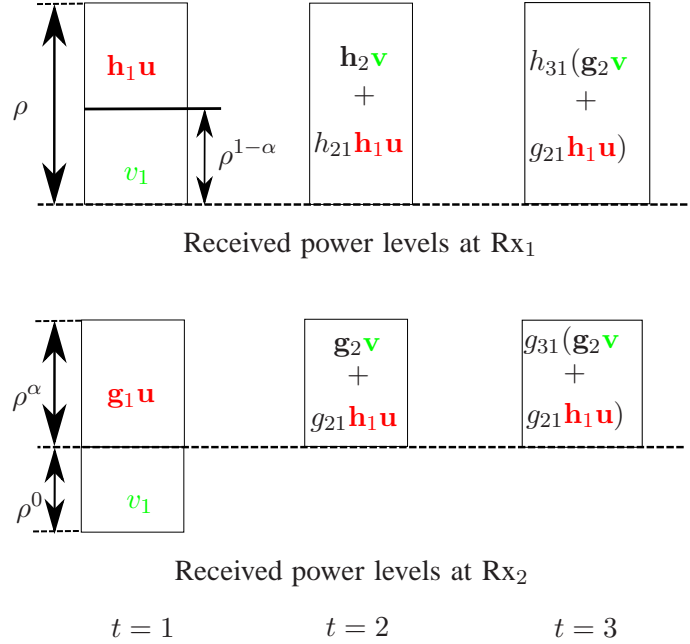}
   \caption{Received power levels at receiver 1 and receiver 2.}
   \psfragscanoff
   \label{power-level}
   \end{figure}
The encoding scheme uses some elements from the compute-and-forward scheme in~\cite{nazar} and the one in Proposition~\ref{prop1}. In this scheme the transmitter wants to send three symbols $(v_1, v_2, v_3)$ to receiver 1 (legitimate receiver) and wishes to conceal them from receiver 2. The coding scheme consists of three phases, each comprising of only one timeslot. In the first phase, the leverage provided by the topology of the network is utilized as follows. The transmitter injects artificial noise from both antennas, where the  output at the receiver 2  is obtained at a lower power level $(\mc O (\rho^\alpha))$ compared to the legitimate receiver $(\mc O (\rho))$. Thus, by reducing the transmission power of the artificial noise to the order of receiver 2  $(\mc O (\rho^\alpha))$, the transmitter can  use the remaining power $(\mc O (\rho^{1-\alpha}))$ to send a confidential symbol to receiver 1. The receiver 2 will receive the confidential symbol embedded in with artificial noise but at noise floor level; and hence, can not decode it. In this phase, as opposed to the scheme in Proposition~\ref{prop1}, the transmitter uses structured codes to send artificial noise, where both receivers are able to compute a unique function of artificial noise. This unique function can be interpreted as a secret key which will be used in the next time slot to secure  information. Let $u_i \in \Lambda$, where $u_i$ denotes the noise codeword chosen from the lattice codebook $\Lambda$, for $i=1,2$.\footnote{For details about lattice codes, the interested reader may refer to \cite{nazar,erez}.} Note that each transmit antenna encodes an independent stream of noise --- no coding across sub channels. After encoding, the transmitter sends structured noise $(u_1, u_2)$ along with a confidential symbol $v_1$ chosen from a Gaussian codebook as
 \begin{eqnarray}
 {\dv{x}}_1 = \left [
 \begin{matrix}
 u_{1}\\
 u_{2}
 \end{matrix} \right ] + \left [
 \begin{matrix}
 v_1 \rho^{-\alpha/2} \\
\phi
 \end{matrix} \right ] \:\:\:\:\:\:
 \end{eqnarray}
 \noindent  The channel input-output relationship is given by
 \begin{subequations}
 \begin{align}
 \label{n1}
 y_1 &=\underbrace{\sqrt{\rho}{{\dv{h}}}_1 {\dv{u}}}_{\mc O (\rho)}+ \underbrace{\sqrt{\rho^{(1-\alpha)}}h_{11}v_1}_{\mc O(\rho^{1-\alpha})}, \\
 \label{n2}
 z_1 &= \underbrace{\sqrt{\rho^{\alpha}}{{\dv{g}}}_1 {\dv{u}}}_{\mc O(\rho^{\alpha})}+ \underbrace{\sqrt{\rho^{0}}g_{11}v_1}_{\mc O(\rho^0)}.
 \end{align}
 \end{subequations}
 At the end of phase 1, each receiver conveys the past CSI to the transmitter. Figure~\ref{power-level} illustrates the received power levels at  receiver 1 and  2, respectively. At the end of phase 1, receiver 2 gets a linear combination of noise codewords along with the confidential symbol $v_1$ at noise floor level; and, thus, can not decode it. Receiver 1 gets the confidential symbol embedded in with a linear combination of noise codewords $\dv{h}_1\dv{u}$. It 
 first re-constructs $\dv{h}_1\dv{u}$ from the channel output $y_1$ by treating $v_1$ as noise, within bounded noise distortion. Since  $\sum_{i=1}^2 a_{ji}u_i \in \Lambda$ where $a_{ji} \in \mb Z$,  $(i,j) = \{1,2\}^2$,  the two receivers can decode the corresponding equations ($\dv{h}_1\dv{u}$, $\dv{g}_1\dv{u}$) as long the computation rate~\cite{nazar} 
\begin{eqnarray}
\label{lattice}
R_\text{comp} &\le & \min \big\{ R(\dv{h}_1, \dv{a}_1:= \dv{h}_1),  R(\dv{g}_1, \dv{a}_1:= \dv{g}_1)\big\} 
\end{eqnarray} 
where   
\begin{eqnarray*}
R(\dv{h}_1, \dv{a}_1:= \dv{h}_1) &=& \log^+ \bigg(\big(||\dv{a}_1||^2+\rho^{(1-\alpha)}||h_{11}||^2 - \frac{\rho ||\dv{h}_1\dv{a}_1||}{1+\rho||\dv{h}_1||^2}\big)^{-1}\bigg)\notag\\ 
R(\dv{g}_1, \dv{a}_1:= \dv{g}_1) &=& \log^+ \bigg(\big(||\dv{a}_1||^2- \frac{\rho^{\alpha} ||\dv{g}_1\dv{a}_1||}{1+\rho^{\alpha}||\dv{g}_1||^2}\big)^{-1}\bigg)
\end{eqnarray*}
is satisfied.\footnote{In this scheme, it is sufficient if receiver 1 can only decode the linear combination of artificial noise. The condition that both receivers decode unique key is only required for the broadcast case.} It can be readily shown from~\eqref{lattice} that, $R_\text{comp}$ yields a GDoF of $\alpha$. After decoding $\dv{h}_1\dv{u}$, the transmitter 
subtracts out the contribution of $\dv{h}_1\dv{u}$ from $y_1$ and decodes $v_1$ through channel inversion. The information transmitted to receiver 1 via symbol $v_1$ is given by
 \begin{eqnarray}
 R_{v_1} &=& I(v_1;y_1|\dv{h}_1\dv{u}) \notag\\
 &=& h(\sqrt{\rho}\dv{h}_1 \dv{u}+\sqrt{\rho^{(1-\alpha)}}h_{11}v_1|\dv{h}_1 \dv{u}) \notag\\
 &=& (1-\alpha)\log(\rho).
 \end{eqnarray}  
 \noindent In the second phase, the transmitter transmits fresh information ($\dv{v}:=[v_2, v_3]^T$) to receiver 1 along with a linear combination of channel output $({{\dv{h}}}_1 {\dv{u}})$ at receiver 1 during the first phase. Since the transmitter already knows $\dv{u}$ and due to the availability of past CSI of receiver 1 (${{\dv{h}}}_1$) in phase 1, it can easily construct ${{\dv{h}}}_1 {\dv{u}}$ and sends
 \begin{eqnarray}
 {\dv{x}}_2 = \left [
 \begin{matrix}
 v_{2}\\
 v_{3}
 \end{matrix} \right ] + \left [
 \begin{matrix}
 {{\dv{h}}}_1 {\dv{u}} \\
 \phi
 \end{matrix} \right ].
 \end{eqnarray}
 \noindent   The channel input-output relationship is given by
 \begin{subequations}
 \begin{align}
 \label{r21}
 y_2 &=\sqrt{\rho}{{\dv{h}}}_2 {\dv{v}}+ \sqrt{\rho}h_{21}{{\dv{h}}}_1 {\dv{u}}, \\
 \label{r22}
 z_2 &=\sqrt{\rho^{\alpha}}{{\dv{g}}}_2 {\dv{v}}+ \sqrt{\rho^{\alpha}} g_{21}{{\dv{h}}}_1 {\dv{u}}.
 \end{align}
 \end{subequations}
 At the end of phase 2, each receiver conveys the past CSI to the transmitter. Since the receiver 1 knows the CSI $({{\dv{h}}}_2)$ and also the channel output $y_1$ from phase 1, it subtracts out the contribution of $\dv{h}_1 \dv {u}$ from the channel output $y_2$, to obtain one equation with two unknowns ($ {\dv{v}}:=[v_2,v_3]^T$). Thus, receiver 1 requires one extra equation to successfully decode the intended variables, being available as interference (side information) at receiver 2.
 
 In the third phase, due to the availability of delayed CSIT, the transmitter can construct the side information $z_2$ at receiver 2 and sends
 \begin{eqnarray}
 {\dv{x}}_3 = \left [
 \begin{matrix}
 {{\dv{g}}}_2 {\dv{v}}+ g_{21}{{\dv{h}}}_1 {\dv{u}}\\
 \phi
 \end{matrix} \right ] .
 \end{eqnarray}
 The channel input-output relationship is given by
 \begin{subequations}
 \begin{align}
 y_3 &= \sqrt{\rho}{h_{31}{\dv{g}}}_2 {\dv{v}} +  \sqrt{\rho}h_{31} g_{21}{{\dv{h}}}_1 {\dv{u}}, \\
 z_3 &=  \sqrt{\rho^{\alpha}}{g_{31}{\dv{g}}}_2 {\dv{v}} +  \sqrt{\rho^{\alpha}} g_{31}g_{21}{{\dv{h}}}_1 {\dv{u}}.
 \end{align}
 \end{subequations}
 At the end of phase 3, by using $y_1$  receiver 1 subtracts out the contribution of ${{\dv{h}}}_1 {\dv{u}}$  from ($y_2,y_3$) and decodes ${\dv{v}}$ through channel inversion. Thus, at the end of three timeslot, 3 symbols are transmitted to receiver 1 which contains $((1-\alpha)+2)\log(\rho)$ bits.

 \vspace{.5em}
 \textit{\underline{Leakage Analysis}.}
 We can write the channel output at the eavesdropper in compact form as
 \iffalse
 \begin{align}
 {\dv{z}}&:=   \underbrace{
 \begin{pmat}[{|}]
 \sqrt{\rho^{0}} {{g}}_{11} & \sqrt{\rho^{\alpha}} {\dv{g}}_{1} & {0} \cr
 0 & \sqrt{\rho^{\alpha}}  {{g}}_{21}{\dv{h}}_{1}  & \sqrt{\rho^{\alpha}} \cr
 0 & \sqrt{\rho^{\alpha}} {{g}}_{31} {{g}}_{21}{\dv{h}}_{1}  & \sqrt{\rho^{\alpha}}  {{g}}_{31}\cr
 \end{pmat} }_{\dv{G}  \:\in\: \mb{C}^{4\times 3}}\left [
 \begin{matrix}
 {v}_1\cr
 {\dv{u}} \cr
 {\dv{g}}_2 {\dv{v}}  \cr
 \end{matrix} \right ],
 \end{align}.
 \fi

 \begin{align}
 {\dv{z}}&:= 
  \left[ \begin{matrix}
   \sqrt{\rho^{0}} {{g}}_{11} & \sqrt{\rho^{\alpha}} {\dv{g}}_{1} & {0} \cr
   0 & \sqrt{\rho^{\alpha}}  {{g}}_{21}{\dv{h}}_{1}  & \sqrt{\rho^{\alpha}} \cr
   0 & \sqrt{\rho^{\alpha}} {{g}}_{31} {{g}}_{21}{\dv{h}}_{1}  & \sqrt{\rho^{\alpha}}  {{g}}_{31}\cr
   \noalign{\vspace{-.5\normalbaselineskip}}
 & \multicolumn{2}{c}{$\upbracefill$}&\cr
 \noalign{\vspace{-.5\normalbaselineskip}}
 &\multicolumn{2}{c}{\dv{G}  \:\in\: \mb{Z}^{3\times 3}}& 
   \end{matrix}\right ] \left [
   \begin{matrix}
   {v}_1\cr
   {\dv{u}} \cr
   {\dv{g}}_2 {\dv{v}}  \cr
   \end{matrix} \right ].
 \end{align}
 \vspace{.5\normalbaselineskip} 
 
 The information rate leaked to receiver 2 is bounded by 
 \begin{align}
 \label{sec-coding}
 I(v_1, \dv{v};\dv{z}|\dv{S}^n) & = I(v_1;\dv{z}|\dv{S}^n)+I(\dv{v};\dv{z}|v_1,\dv{S}^n)\notag\\
 & \overset{(a)}{=} I(v_1;z_1|\dv{S}^n)+I(\dv{v};\dv{z}|v_1,\dv{S}^n)\notag\\
  &\le o(\log (\rho))+ I(\dv{g}_2\dv{v}, \dv{u};\dv{z}|v_1,\dv{S}^n)- I(\dv{u};\dv{z}|\dv{g}_2\dv{v}, v_1,\dv{S}^n)\notag\\
 & = o(\log (\rho))+\text{rank} (\dv{G}). \log(\rho^\alpha)-2\alpha \log(\rho) \notag\\
 & = o(\log (\rho))+2\alpha \log(\rho)-2\alpha \log(\rho) \notag\\
 & =  o(\log (\rho)).
 \end{align}
 \noindent where $(a)$ follows due to the Independence of  $v_1$ and  $(z_2, z_3)$. From the above analysis, it can be easily seen that $3$ symbols are transmitted to  receiver 1 over a total of $3$ time slots, yielding $\frac{2+(1-\alpha)}{3}$ GSDoF at receiver 1. 
  \end{IEEEproof}
 \vspace{.5em}
 \begin{remark}
 From~\eqref{sec-coding}, the information leakage to the eavesdropper is 
 \begin{eqnarray}
 \limsup_{n \rightarrow \infty} \frac{I(W_1;z^n|\dv{S}^n)}{n} = o (\log (\rho)).
 \end{eqnarray}
 Next, we strengthen the scheme in Proposition~\ref{prop1-new} by combining it with random coding argument used in Wyner's wiretap coding~\cite{wyner} such that~\eqref{sec-constraint-2} is satisfied. We consider an equivalent $\tilde{n}$-block transmission model, where the total duration of the coding scheme in Proposition~\ref{prop1-new} denotes the block length. Let $\tilde {\dv{v}} := (v_1, v_2, v_3)$ denotes the input in each block, $ \tilde{\dv{y}} := (y_1,y_2, y_3)$ denotes the channel output at receiver 1 and $\tilde{\dv{z}}  := (z_1, z_2, z_3)$ denotes the channel output at receiver 2, where the inputs $\tilde {\dv{v}}$ are chosen independently from state sequence $\tilde{\dv{S}}$. This resulting model reduces to the Wyner's wiretap setup~\cite{wyner} where equivocation rate is given by 
 \begin{eqnarray*}
 R_e =  I(\tilde{\dv{v}};\tilde{\dv{y}}|\tilde{\dv{S}}) - I(\tilde{\dv{v}};\tilde{\dv{z}}|\tilde{\dv{S}})
 \end{eqnarray*}
 which satisfies the perfect secrecy criteria of 
  \begin{eqnarray*}
 \limsup_{n \rightarrow \infty} \frac{I(W_1;\tilde{\dv{z}}^n|\tilde{\dv{S}}^n)}{n} = 0.
 \end{eqnarray*} 
 It is worth noting that, by using similar arguments, we can strengthen the security of all schemes in this work which fulfills the perfect secrecy criteria~\eqref{sec-constraint-1} and~\eqref{sec-constraint-2}. 
 \end{remark}
 \vspace{.5em}

\subsection{Coding scheme with symmetric alternating topology}
 \begin{figure}
\psfragscanon
\centering
\psfrag{b}[c][c]{\hspace{-.5em} $\frac{2}{3}$}
\psfrag{a}[c][c]{  $\frac{3+\alpha}{6}$}
\psfrag{x}[c][c]{ $d_1(\alpha)$}
\psfrag{y}[c][c]{\hspace{-1em} $d_2(\alpha)$}
\psfrag{c}[c][c]{$\frac{3+\alpha}{6}$}
\psfrag{d}{\hspace{.5em}$\frac{2}{3}$}
\psfrag{f}{$( \frac{1}{2},\frac{1}{2})$}
\psfrag{g}{Outer bound~\eqref{outer-bound}}
\psfrag{h}{Inner bound~\eqref{theorem-sdof-sym}} 
\includegraphics[width=.6\linewidth]{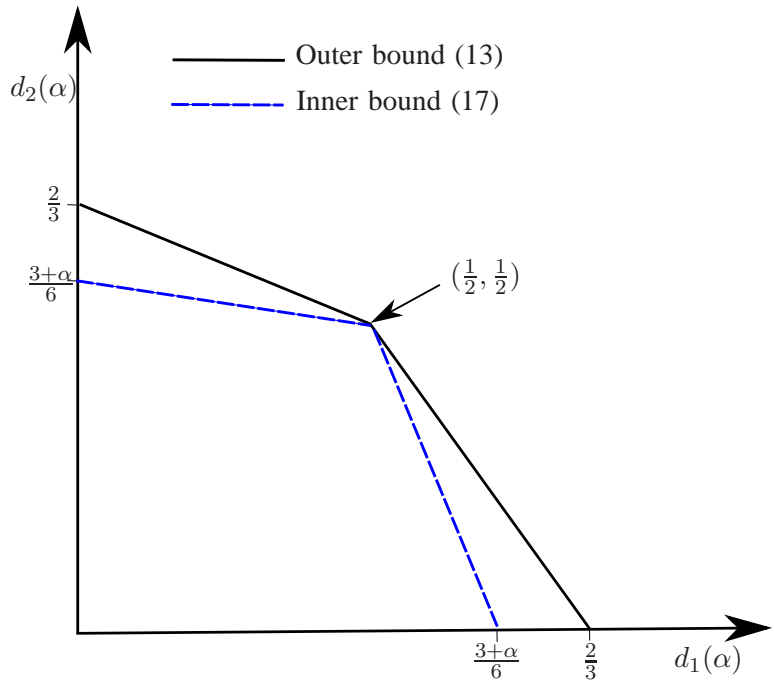}
\caption{Bounds on the GSDoF region of $(2,1,1)$--MISO broadcast channel with symmetric  alternating topology $(\lambda_{1 \alpha},\lambda_{\alpha 1})=( \frac{1}{2},\frac{1}{2})$.}
\psfragscanoff
\label{R-Comp-sym-sec-psfrag}
\end{figure} 

\noindent The following proposition provides the sum GSDoF of the MISO broadcast channel with symmetric alternating topology  $(\lambda_{1 \alpha},\lambda_{\alpha 1})= (\frac{1}{2}, \frac{1}{2})$.
\vspace{.5em} 
\begin{proposition}
\label{prop5}
An inner bound on the  GSDoF region $\mc C_\text{GSDoF}(\lambda_{A_1 A_2})$ of the two user $(2,1,1)$--MISO broadcast channel with delayed CSIT and symmetric alternating topology  $(\lambda_{1 \alpha}=\lambda_{\alpha 1}= \frac{1}{2})$ is given by the set of all non-negative pairs $(d_1,d_2)$ satisfying
\begin{subequations}
\label{theorem-sdof-sym-new}
\begin{align}
\label{miso-bc-outer-a}
3d_1+\alpha d_2 &\le \frac{3+ \alpha}{2} \\
\label{miso-bc-outer-b}
 \alpha d_1+3d_2 &\le \frac{3+ \alpha}{2}. 
\end{align}
\end{subequations}
\end{proposition}
\vspace{.5em} 
\begin{IEEEproof}
 As shown in Figure~\ref{R-Comp-sym-sec-psfrag}, it is sufficient to prove the GSDoF pairs $(\frac{3+\alpha}{6},0)$,  $(0,\frac{3+\alpha}{6})$ and $(\frac{1}{2},\frac{1}{2})$, since the entire region~\eqref{theorem-sdof-sym-new} can then be achieved by time sharing. The GSDoF pairs $(\frac{3+\alpha}{6},0)$ and $(0,\frac{3+\alpha}{6})$ are achievable by using the combination of coding schemes developed in Proposition~\ref{prop1-new} and in Section~\ref{fixed-topology-3}, equal fractions of time. The achievability of the  point $(\frac{1}{2},\frac{1}{2})$ is provided in Appendix~\ref{app-6}.
 \end{IEEEproof}

\subsection{GDoF region with fixed topology}
We now consider the MISO broadcast channel studied in Section~\ref{special-base} with no secrecy constraints and characterize the GDoF region. The following theorem provides the GDoF region of the MISO broadcast channel with fixed topology  $(\lambda_{1 \alpha}=1)$.
\vspace{.5em} 
\begin{theorem}
\label{prop3}
The GDoF region $\mc C_\text{GDoF}(\lambda_{A_1 A_2})$ of the two user $(2,1,1)$--MISO broadcast channel with delayed CSIT and fixed topology  $(\lambda_{1 \alpha}=1)$ is given by the set of all non-negative pairs $(d_1,d_2)$ satisfying
\begin{subequations}
\label{theorem-sdof-miso-bc-outer-bound-eqautions}
\begin{align}
\label{m1}
d_1 &\le 1 \\
\label{m2}
d_2 &\le \alpha \\
\label{m3}
2d_1+d_2 &\le 2 \\
\label{m4}
d_1+2d_2 &\le 1+\alpha. 
\end{align}
\end{subequations}
\end{theorem}
\vspace{.5em} 
\begin{IEEEproof}
The converse immediately follows from the outer bound established in ~\cite[eq. 30-31]{CEJ14}. As seen in Figure~\ref{R-Comp-asym-psfrag}, it suffices to prove that following GDoF pairs  $(0,\alpha)$,  $(1,0)$, $(1-\alpha,\alpha)$ and $(1-\frac{\alpha}{3},\frac{2\alpha}{3})$ are achievable. The GDoF pairs $(0,\alpha)$, and  $(1,0)$ are readily achievable even without utilizing the delayed CSI, by transmitting information to only one receiver, since the equivalent model reduces to a point-to-point channel with two transmit antennas and a single receive antenna.
\begin{figure}
\psfragscanon
\centering
\psfrag{a}{\hspace{.5em}$1$}
\psfrag{b}{\hspace{.5em} $\alpha$}
\psfrag{x}{ $d_1(\alpha)$}
\psfrag{y}{\hspace{-1em} $d_2(\alpha)$}
\psfrag{c}{$(1-\frac{\alpha}{3}, \frac{2\alpha}{3})$}
\psfrag{d}{$(1- \alpha ,  \alpha )$}
\includegraphics[width=.6\linewidth]{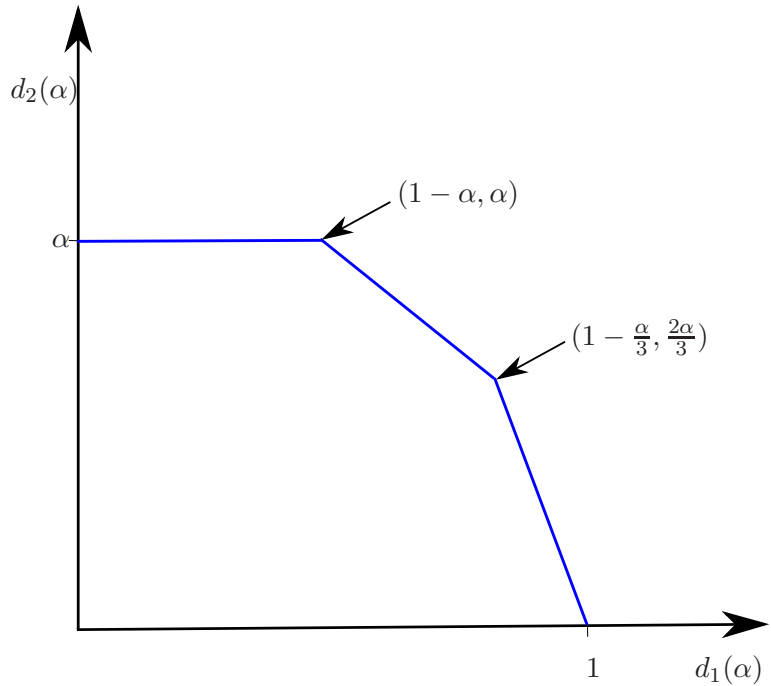}
\caption{GDoF region~\eqref{theorem-sdof-miso-bc-outer-bound-eqautions} of $(2,1,1)$--MISO broadcast channel, where the link to receiver 2 is weaker than receiver 1 ($\lambda_{1 \alpha}:=1$).}
\psfragscanoff
\label{R-Comp-asym-psfrag}
\end{figure} 

The achievability of the GDoF pair $(1-\alpha,\alpha)$ follows by sending one symbol $(v)$ to receiver 1 and one symbol  $(w)$ to receiver 2 as follows
\begin{eqnarray}
{\dv{x}}_1 = \left [
\begin{matrix}
w+v \rho^{-\alpha/2}\\\
\phi
\end{matrix} \right ] 
\end{eqnarray}
\noindent  where $\mathbb{E}[||w||^2]\doteq 1$ and  $\mathbb{E}[||v^2||] \doteq 1 $.
The channel input-output relationship is given by
\begin{subequations}
\begin{align}
\label{nd1}
y_1 &=\underbrace{\sqrt{\rho}{{{h}}}_{11} w}_{\mc O (\rho)}+ \underbrace{\sqrt{\rho^{(1-\alpha)}}h_{11}v}_{\mc O(\rho^{1-\alpha})}, \\
\label{nd2}
z_1 &= \underbrace{\sqrt{\rho^{\alpha}}{g}_{11} w}_{\mc O(\rho^{\alpha})}+ \underbrace{\sqrt{\rho^{0}}g_{11}v}_{\mc O(\rho^0)}.
\end{align}
\end{subequations}
At the end of transmission, receiver 1 first re-constructs $h_{11}w$ from the channel output $y_1$ by treating $v$ as noise, within bounded noise distortion. Afterwards, it  decodes $v$ by subtracting out the contribution of $w$ from $y_1$.
The information transmitted to receiver 1 via symbol $v$ is given by
 \begin{eqnarray}
 R_{v_1} &=& I(v ;y_1|h_{11}w) \notag\\
 &=& (1-\alpha)\log(\rho)
\end{eqnarray}
which yields a GDoF of $1-\alpha$ at receiver 1.
Using similar reasoning and algebra, it can be readily shown that $\alpha$ GDoF is achievable at receiver 2. 

The achievability of the point $(1-\frac{\alpha}{3},\frac{2\alpha}{3})$ follows by specializing the encoding scheme that we establish in Proposition~\ref{prop2} by removing the secrecy constraints. The proof of the achievability is relegated to Appendix~\ref{app-5}.
\end{IEEEproof}
\iffalse
 \vspace{.5em}
\begin{remark}

\end{remark}
 \vspace{.5em}
 \fi

\section{Numerical Examples}\label{V}
\begin{figure}
\psfragscanon
\centering 
\psfrag{a}{\hspace{.5em}$1$}
\psfrag{b}{\hspace{.5em} $\alpha$}
\psfrag{x}{ $d_1(\alpha)$}
\psfrag{y}{\hspace{-1em} $d_2(\alpha)$}
\psfrag{c}{$\big(\frac{2}{2+\alpha}, \frac{\alpha(1+\alpha)}{2+\alpha}\big)$}
\psfrag{d}{$(1- \alpha ,  \alpha )$}
\psfrag{e}{$ \frac{2\alpha}{3}$}
\psfrag{f}{\hspace{.25em}$\frac{2}{3}$}
\psfrag{i}{\hspace{-2em}$\big(\frac{2}{3+\alpha}, \frac{\alpha(1+\alpha)}{3+\alpha}\big)$}
\psfrag{g}{GDoF region~\cite[Propostion 2]{CEJ14}}
\psfrag{h}{GSDoF region~\eqref{prop-2-eqautions}} 
\includegraphics[width=.6\linewidth]{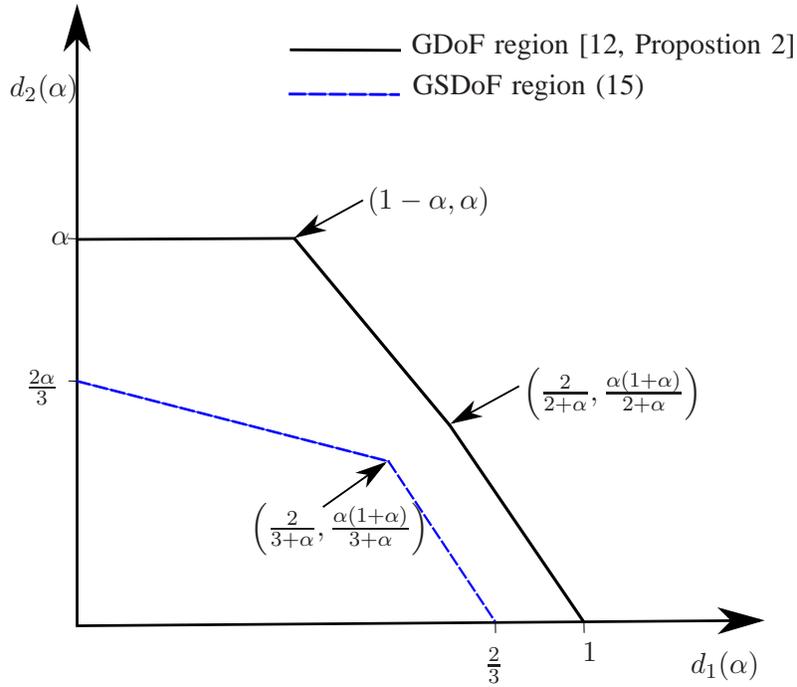}
\caption{GDoF and GSDoF regions of MISO broadcast channel with fixed topology $(\lambda_{1\alpha})$.}
\psfragscanoff
\label{comp-optimal}
\end{figure}  
\begin{figure}
\psfragscanon
\centering
\psfrag{a}{\hspace{.5em}$0$}
\psfrag{b}{ $1$}
\psfrag{x}{ $\alpha$}
\psfrag{y}{$d_1+d_2$}
\psfrag{c}{$ \frac{1}{2}$}
\psfrag{d}{$\frac{3}{4}$}
\psfrag{e}{$1$}
\psfrag{f}{$ \frac{4}{3}$}
\psfrag{o}{$ \frac{2}{3}$}
\psfrag{i}{$(1-\frac{\alpha}{2}, \frac{\alpha}{2})$}
\psfrag{g}{Optimal sum GDoF with symmetric alternating topology\cite[eq. (12)]{CEJ14}}
\psfrag{k}{Sum GDoF with fixed topology\cite[eq. (11)]{CEJ14}}
\psfrag{l}{Sum GSDoF with  symmetric alternating topology~\eqref{theorem-sdof-sym}}
\psfrag{h}{Sum GSDoF with fixed topology~\eqref{prop-2-eqautions}} 
\psfrag{i}{Trivial lower bound on sum GSDoF with fixed topology~\eqref{trivial-sdof-miso-bc-outer-bound-eqautions}} 
\includegraphics[width=.6\linewidth]{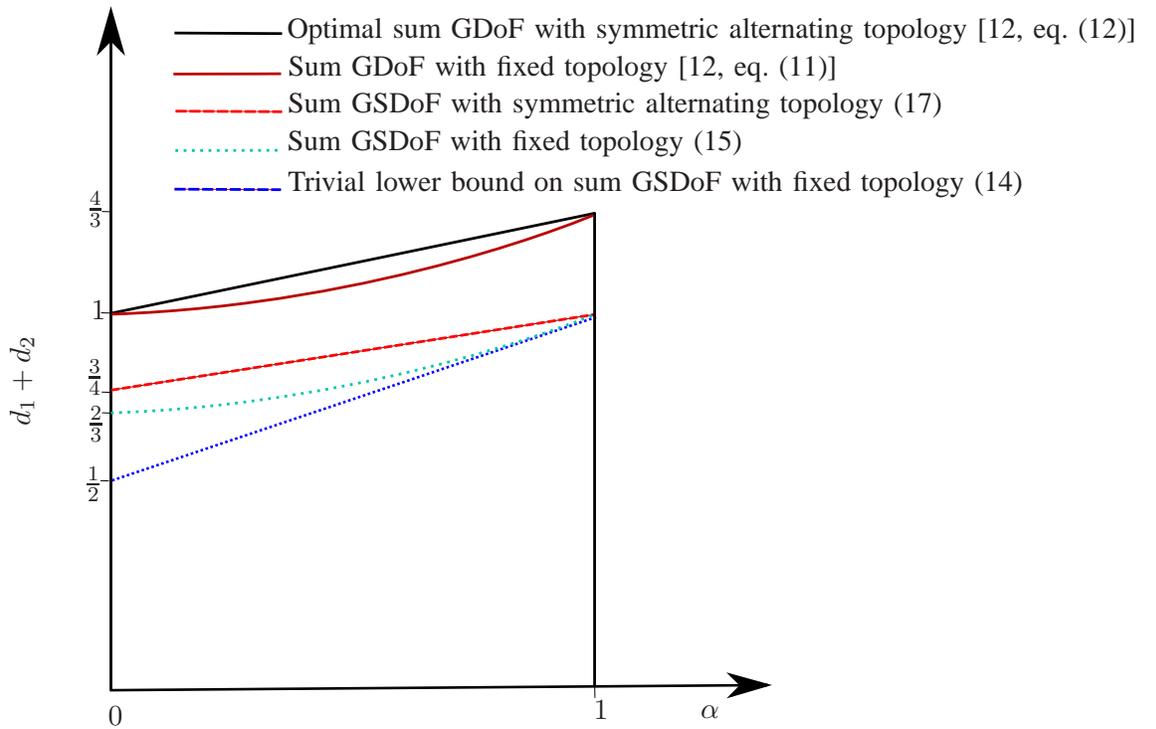}
\caption{Sum GDoF and GSDoF of MISO broadcast channel with different topology consideration.}
\psfragscanoff
\label{sum-comp-optimal}
\end{figure}  
In this section, we provide some numerical examples to illustrate the results presented in the previous sections. Figure~\ref{comp-optimal} shows an inner bound on the GDoF region for the MISO broadcast channel with delayed CSIT and fixed network topology given by~\cite[Propostion 2]{CEJ14}. In order to illustrate the loss incurred in GDoF region, we also plot the inner bound on GSDoF region of a similar model given by  Proposition~\ref{prop2}. As we have mentioned before, albeit the optimality of the inner bound of Proposition~\ref{prop2} is still to be shown, the visible gap between the two regions illustrates the loss in terms of GDoF due to  secrecy constraints.

This loss is also reflected in Figure~\ref{sum-comp-optimal}, where we plot the generalized sum (or total) DoF of MISO broadcast channel and delayed CSIT with fixed and symmetric alternating network topology given by~\cite[eq. (11)]{CEJ14} and~\cite[eq. (12)]{CEJ14} as a function of network topology parameter $\alpha$. Figure~\ref{sum-comp-optimal} also shows the  sum GSDoF of MISO broadcast channel with fixed and symmetric alternating topology given by Proposition~\ref{prop2} and~\ref{prop4}, respectively. As a reference, we also plot a lower bound on sum GSDoF obtained by straightforward adaptation of Yang \ti{et al.} scheme given by~\eqref{trivial-sdof-miso-bc-outer-bound-eqautions}. It can be easily seen from the figure that, as the parameter $\alpha$ approaches 1, i.e., all links have equal strength, the inner bounds  in Proposition~\ref{prop2} and~\ref{prop4} recover the sum SDoF of a similar model in the absence of network topology~\cite{YKPS11}. We note that this equivalence also holds for models without secrecy constraints. In particular, as shown in the Figure~\ref{sum-comp-optimal} as  $\alpha$ approaches $1$, one can recover the optimal sum DoF of Maddah-Ali-Tse (MAT) scheme~\cite{M-AT12}.
 
\section{Conclusion}\label{conclusion}
 
In this paper, we study the GSDoF of a MISO broadcast channel. We assume that perfect CSI is available at the receivers and each receiver only conveys the past CSI to the transmitter. In addition to this, links connecting both receivers may have unequal strength, statistically. We first consider the MISO wiretap channel and establish bounds on GSDoF. For the case in which the legitimate receiver is comparatively stronger than the eavesdropper, under certain conditions, the lower and upper bounds agree, and so, we characterize the GSDoF. Next, we extend this model to the broadcast setting and establish bounds on GSDoF region. The coding scheme is based on an appropriate extension of noise injection scheme~\cite{YKPS11}, where the transmitter utilizes the knowledge of network topology and past CSI in a non-trivial manner. Furthermore, we specialize our result to the model with no secrecy constraints and   characterize the DoF region for the topology state in which one receiver is stronger than the other. The results establish in this work highlight the interplay between network topology and CSI, and sheds light on how to efficiently utilize both resources in securing information.
 
\appendices
\section{Proof of Lemma~\ref{lemma}}
\label{app-1}
Before proceeding to the proof of Lemma~\ref{lemma}, for completeness, we first introduce a property~\cite[Lemma 4]{vaze_int} which is used to establish the results in this work. 

\noindent Recall that the channel output at receiver 2 is given by 
\begin{eqnarray}
\label{n11}
{z}_t = \sqrt{{\rho}^{A_{2,t}}} {\dv{g}}_t \dv x_t + {n}_{2t}.
\end{eqnarray}
Now, lets consider an artificial channel $\tilde {z}_i$ at receiver 2, such that the channel input-output relationship at $i$-th time instant is  
\begin{eqnarray}
\label{n12}
\tilde {z}_t = \sqrt{{\rho}^{A_{2,t}}} \tilde {\dv{g}}_t \dv x_t + \tilde{n}_{2t}
\end{eqnarray}
where $\tilde {\dv{g}}_t$ and $\tilde{n}_{2t}$ are independent form eachother and identically distributed as ${\dv{g}}_t$ and ${n}_{2t}$, respectively.   Let $\lambda_{\dv{g}_t}$ denotes the probability distribution from which, $\dv{g}_t$ and $\tilde {\dv{g}}_t$  are independent and identically drawn. Let $\dv{S}^n:=\{\dv{g}_t ,\tilde {\dv{g}}_t\}_{t=1}^n$. 
\vspace{.5em}
\begin{property}
The channel output symmetry states that 
 \setlength{\arraycolsep}{0.1em}
\begin{eqnarray}
\label{sym-eq}
h(z_t|z^{t-1},\dv{S}^n)&=&  h(\tilde {z}_t|z^{t-1}, \dv{S}^n ) .
\end{eqnarray}
\end{property}
\vspace{.5em}

\begin{proof}
We begin the proof as follows.
\begin{eqnarray}
h(z_t|z^{t-1},\dv{S}^n)&=&   h(z_t|z^{t-1}, \dv{g}_t,\tilde {\dv{g}}_t,\dv{S}^n\setminus\dv{S}_t) \notag\\ 
\label{tdenttty2}
&=& \mb {E}_{\lambda_{ \dv{g}_t}} [h(\sqrt{{\rho}^{A_{2,t}}}\dv{g}\dv{x}_t+n_{2t}|z^{t-1}, \dv{g}_t=\dv{g},\tilde {\dv{g}}_t, \dv{S}^n\setminus\dv{S}_t )]\notag\\ 
&\overset{(a)}{=}& \mb {E}_{\lambda_{ \dv{g}_t}} [h(\sqrt{{\rho}^{A_{2,t}}}\dv{g}\dv{x}_t+\tilde {n}_{2t}|z^{t-1},\dv{S}^n\setminus\dv{S}_t )]\notag\\ 
&\overset{(b)}{=}& \mb {E}_{\lambda_{ \dv{g}_t}} [h(\sqrt{{\rho}^{A_{2,t}}}\dv{g}\dv{x}_t+ \tilde{n}_{2t}|z^{t-1}, \tilde {\dv{g}}_t=\dv{g},\dv{g}_t, \dv{S}^n\setminus\dv{S}_t )]\notag\\ 
\label{tdenttty4}
&\overset{(c)}{=}& \mb {E}_{\lambda_{ \dv{g}_t}} [h(\tilde {z}_t|z^{t-1}, \dv{g}_t,\tilde {\dv{g}}_t=\dv{g}, \dv{S}^n\setminus\dv{S}_t )]\notag\\ 
&{=}& h(\tilde {z}_t|z^{t-1}, \dv{S}^n ) 
\end{eqnarray}
\noindent where $(a)$ follows because $n_{2t}$ and $\tilde {n}_{2t}$ are independent from $(\dv{x}_t,\dv{g}_t,\tilde {\dv{g}}_t)$ and have same statistics, $(b)$ follows since $\dv{g}_t$ and $\tilde {\dv{g}}_t$ belongs to  $\lambda_{\dv{g}_t}$ and have the same alphabet set; and $(c)$ follows due to the independence of $\dv{x}_t$ and $(\dv{g}_t, \tilde {\dv{g}}_t)$.
\end{proof}
\vspace{.5em}

\noindent We now provide the proof of~\eqref{l1} and \eqref{l3}; due to the symmetry the rest of the inequalities follow straightforwardly. For convenience, we first denote the channel output as
\begin{align}
 z^n := (z^n_{11}, z^n_{1\alpha}, z^n_{\alpha 1}, z^n_{\alpha\alpha}) \notag
\end{align}
where $ z^n_{A_1A_2}$ denotes the part of channel output, when $(A_1,A_2) \in \{1,\alpha\}^2$ channel state occurs.

We begin the proof by defining an auxiliary random variable $u_t$, such that, given  $(z_t, \tilde{z}_t, u_t, \dv{S}_t)$, $y_t$ is recovered fully. From~\eqref{n11} and~\eqref{n12}, the channel outputs at receiver 2 and artificial receiver are given by 
\begin{eqnarray}
\label{multi-n}
\left [
\begin{matrix}
z_t \\
\tilde {z}_t
\end{matrix} \right]
= \sqrt{\rho^{ A_{2,t}}} \left [
\begin{matrix}
\dv{g}_t \\
\tilde {\dv{g}}_t
\end{matrix} \right]\dv {x}_t + \left[
\begin{matrix}
n_{2t} \\
\tilde {n}_{2t}
\end{matrix} \right].  
\end{eqnarray}
Then,  scaling~\eqref{multi-n} with $\sqrt{\rho^{(A_{1,t}-A_{2,t})}} \left [
\begin{smallmatrix}
\dv{g}_t \\
\dv{h}_t
\end{smallmatrix} \right]\left [
\begin{smallmatrix}
\dv{g}_t \\
\tilde {\dv{g}}_t
\end{smallmatrix} \right]^{-1}$, we get
\begin{eqnarray}
\label{equation-main}
\iffalse
\rho^{\frac{(A_{1,t}-A_{2,t})}{2}} \left [
\begin{matrix}
\dv{g}_t \\
\dv{h}_t
\end{matrix} \right]\left [
\begin{matrix}
\dv{g}_t \\
\tilde {\dv{g}}_t
\end{matrix} \right]^{-1}\left [
\begin{matrix}
z_t \\
\tilde{z}_t
\end{matrix} \right] =\fi &=& \underbrace{\sqrt{\rho^{ A_{1,t} }}\left [
\begin{matrix}
\dv{g}_t \\
\dv{h}_t
\end{matrix} \right]\dv{x}_t+ \left [
\begin{matrix}
0 \\
n_{1t}
\end{matrix} \right]}_{= \: \left [
\begin{matrix}
\phi_1 \\
y_t
\end{matrix} \right]  } + \underbrace{\sqrt{\rho^{(A_{1,t}-A_{2,t})}} \left [
\begin{matrix}
\dv{g}_t \\
\dv{h}_t
\end{matrix} \right]\left [
\begin{matrix}
\dv{g}_t \\
\tilde {\dv{g}}_t
\end{matrix} \right]^{-1}\left [
\begin{matrix}
n_{2t} \\
\tilde {n}_{2t}
\end{matrix} \right]+ \left [
\begin{matrix}
0 \\
-n_{1t}
\end{matrix} \right]}_{:=\: \left [
\begin{matrix}
\phi_2 \\
u_t
\end{matrix} \right]} 
\end{eqnarray}
where $\mathbb{E}[||u_t||^2] \dot{=} \rho^{(A_{1,t}-A_{2,t})^+}$. Then, it can be easily seen from~\eqref{equation-main} that, given $\dv{S}^n$, by subtracting the contribution of $u_t$ from $(z_t, \tilde{z}_t)$, suffices to construct $y_t$.

Now, we proceed as follows.  
\begin{eqnarray}
\label{z1}
h(z^n|\dv{S}^n)=  \sum_{t=1}^n h(z _t|z^{t-1},\dv{S}^n)\\
\label{z2}
h(z^n|\dv{S}^n)=  \sum_{t=1}^n h(\tilde {z}_t|z^{t-1},\dv{S}^n) 
\end{eqnarray}
where~\eqref{z2} follows due to the property of channel output symmetry~\eqref{sym-eq}. Then, by combining~\eqref{z1} and \eqref{z2}, we get
\begin{align}
2 h(z^n|\dv{S}^n)&= \sum_{t=1}^n h(z _t|z^{t-1},\dv{S}^n)+ h(\tilde {z}_t|z^{t-1},\dv{S}^n) \notag\\ 
&\overset{(d)}{\geq} \sum_{t=1}^n h(z _t,\tilde {z}_t |z^{t-1},\dv{S}^n) \notag\\ 
&= \sum_{t=1}^n h(z _t,\tilde {z}_t, y_t, u_t |z^{t-1},\dv{S}^n) -  h(y_t,u_t |z^n,\tilde {z}_t,\dv{S}^n)\notag\\
&= \sum_{t=1}^n h(z _t, y_t |z^{t-1},\dv{S}^n)+h(\tilde {z}_t,u_t|z^n,y_t,\dv{S}^n) -h(u_t |z^n, \tilde {z}_t,\dv{S}^n)-  h(y_t |z^n,\tilde {z}_t,u_t,\dv{S}^n)\notag\\ 
&= \sum_{t=1}^n h(z _t, y_t |z^{t-1},\dv{S}^n)+ h(\tilde {z}_t|z^n,y_t,\dv{S}^n) +\underbrace{h(u_t|z^n,\tilde{z}_t, y_t,\dv{S}^n)}_{=0}   -h(u_t |z^n, \tilde {z}_t,\dv{S}^n)\notag\\&\quad-  \underbrace{h(y_t |z^n,\tilde {z}_t,u_t,\dv{S}^n)}_{=0}\notag\\ 
&\overset{(e)}{=} \sum_{t=1}^n h(z _t, y_t |z^{t-1},\dv{S}^n)+\underbrace{h(\tilde {z}_t|z^n,y_t,\dv{S}^n)}_{=  n o(\log(\rho))} -h(u_t |z^n, \tilde {z}_t,\dv{S}^n) \notag\\ 
&\overset{(f)}{\geq} \sum_{t=1}^n h(z _t, y_t |z^{t-1},\dv{S}^n) -h(u_t |\dv{S}^n)+n o(\log(\rho))\notag\\   
&\overset{(g)}\geq \sum_{t=1}^n h(z _t, y_t |z^{t-1},y^{t-1},\dv{S}^n)-h(u_t |\dv{S}^n)+n o(\log(\rho))\notag\\ 
&\overset{(h)}=  h(z^n, y^n | \dv{S}^n)-h(u^n |\dv{S}^n)+n o(\log(\rho)) \notag\\ 
&=  h(z^n, y^n | \dv{S}^n)-h(u_{11}^n, u_{1\alpha}^n, u_{\alpha1}^n, u_{\alpha\alpha}^n |\dv{S}^n) +n o(\log(\rho))\notag\\ 
\label{last-term}
&\overset{(i)}{\geq}  h(z^n, y^n | \dv{S}^n) - \sum_{{A_1,A_2}\in(1,\alpha)^2}n \lambda_{A_1 A_2} \log(\rho^{(A_1-A_2)^+})+n o(\log(\rho))\notag\\
&=  h(z^n, y^n | \dv{S}^n) - n \lambda_{1\alpha}( 1-\alpha)  \log(\rho)+n o(\log(\rho))
\end{align}

\noindent where $(d)$, $(g)$ and $(i)$ follow form the fact that conditioning reduces entropy, $(e)$ follows because $y_t$ and $u_t$ can be fully recovered from  $(z_t,\tilde{z}_t,u_t, \dv{S}^n)$ and $(z_t,\tilde{z}_t,y_t, \dv{S}^n)$, respectively, $(f)$ follows because $\tilde {z}_t$ is constructed within bounded noise distortion from $(z_t,y_t,\dv{S}^n)$; and, $(h)$ follows due to the independence of $u_t$ and $u^{t-1}$.

We can also bound~\eqref{last-term} as follows
\begin{align}
2 h(z^n|\dv{S}^n)+ n\lambda_{1\alpha}(1-\alpha)\log(\rho)
& \geq  h(z^n, y^n | \dv{S}^n)+n o(\log(\rho))\notag\\
& \geq  h(y^n | \dv{S}^n)+n o(\log(\rho)).
\end{align}
This concludes the proof.

\section{Proof of Theorem~\ref{theorem-sdof-wt}}
\label{proof-sdof-wt}
\vspace{.2em}
\iffalse The converse uses elements from the proof established earlier in the context of  wiretap channel with delayed CSIT \cite{YKPS11} and also, uses properties of channel local output symmetry in Lemma~\ref{lemma}.\fi  We now provide the proof of Theorem~\ref{theorem-sdof-wt}. We  begin the proof as follows.
\begin{align}
 nR_{e}&= H(W|z^n, \dv{S}^n ) \notag\\
&= H(W |\dv{S}^n) - I(W; z^n | \dv{S}^n )\notag\\
&= I(W; y^n | \dv{S}^n)+H(W | y^n, \dv{S}^n)- I(W; z^n | \dv{S}^n)\notag\\
\label{mid-wt}
&\overset{(a)}{\le} I(W; y^n | \dv{S}^n)- I(W; z^n| \dv{S}^n)+n\epsilon_{n}\\
\label{b1-wt}
& \le I(W; y^n, z^n | \dv{S}^n)- I(W; z^n| \dv{S}^n)+n\epsilon_{n} \notag\\
& = h(y^n, z^n | \dv{S}^n)-  h(y^n | W, z^n, \dv{S}^n)- h(z^n| \dv{S}^n)+n\epsilon_{n} \notag\\
& \le h(y^n, z^n | \dv{S}^n)- h(z^n| \dv{S}^n)-\underbrace{h(y^n | W, \dv{x}^n, z^n, \dv{S}^n)}_{ \ge no(\log(\rho))}  +n\epsilon_{n} \notag\\
&\overset{(b)}{\le} h(z^n | \dv{S}^n)+n\lambda_{1\alpha}(1-\alpha)\log(\rho)+n\epsilon_{n}
\end{align}
\noindent where $\epsilon_n \rightarrow 0$ as $n \rightarrow \infty$;
$(a)$ follows from Fano's inequality, $(b)$ follows because $(y^n)$ can be obtained within noise distortion form $(\dv{x}^n$, $\dv{S}^n$), and using \eqref{l1}.

We can also bound $R_e$ as follows. From~\eqref{mid-wt}, we get
\begin{align}
\label{b2-wt}
 nR_{e}&\le  I(W; y^n | \dv{S}^n)- I(W; z^n| \dv{S}^n)+n\epsilon_{n} \notag\\
&\overset{(c)}{\le} h(y^n | \dv{S}^n) - \frac{1}{2}h(z^n |W, \dv{S}^n)+\frac{n\lambda_{\alpha 1}(1-\alpha)}{2}\log(\rho) - h(z^n |\dv{S}^n) +  h(z^n| W, \dv{S}^n)+n\epsilon_{n} \notag\\
&\overset{(d)}{\le} h(y^n | \dv{S}^n) - \frac{1}{2}h(z^n |\dv{S}^n)+\frac{n\lambda_{\alpha 1}(1-\alpha)}{2}\log(\rho)+n\epsilon_{n}
\end{align}
\noindent where $(c)$ follows from~\eqref{l4} and $(d)$ follows from the fact that conditioning reduces entropy.

Then combining these two upper bounds in \eqref{b1-wt} and \eqref{b2-wt}, we get
\begin{align}
\label{wt-opt-1}
nR_{e}&\le \min\big\{h(z^n | \dv{S}^n)+n\lambda_{1 \alpha}(1-\alpha)\log(\rho), h(y^n | \dv{S}^n) -\frac{1}{2}h(z^n |\dv{S}^n)+\frac{n\lambda_{\alpha 1}(1-\alpha)}{2}\log(\rho)\big\} +n\epsilon_{n}\\
&\overset{(e)}{\le} \max_{h(y^n)}\frac{2}{3}h(y^n | \dv{S}^n) + \frac{n(1-\alpha)(\lambda_{1 \alpha}+\lambda_{\alpha 1})}{3}\log(\rho)+n\epsilon_{n}\notag\\
&\overset{(f)}{\le} \max_{h(y^n)} \frac{2}{3}\big(h(y^n_{11} | \dv{S}^n)+ h(y^n_{1\alpha}|\dv{S}^n)+h( y^n_{\alpha 1}| \dv{S}^n)+h(y^n_{\alpha\alpha}| \dv{S}^n)\big) + \frac{n(1-\alpha)(\lambda_{1 \alpha}+\lambda_{\alpha 1})}{3}\log(\rho)+n\epsilon_{n}\notag\\
\label{wt-opt-2}
&\le  \frac{(3-\alpha)\lambda_{1 \alpha}+2(\lambda_{11}+\alpha\lambda_{\alpha\alpha})+(1+\alpha)\lambda_{\alpha 1}}{3}n\log(\rho)+n\epsilon_{n} 
\end{align}
\noindent where $(e)$ follows by maximizing~\eqref{wt-opt-1} with respect to $h(z^n| \dv{S}^n)$, and $(f)$ follows from the fact that conditioning reduces entropy. Then, dividing both sides by $n\log(\rho)$ and  taking $\lim \rho \rightarrow \infty$ and $\lim n \rightarrow \infty$, in~\eqref{wt-opt-2}, we get~\eqref{eq-sdof-wt}.

This concludes the proof.

\section{Proof of Theorem~\ref{theorem-sdof-bc}}
\label{proof-sdof-bc}
\vspace{.2em}
The converse follows by generalizing the proof established in Theorem~\ref{theorem-sdof-wt} in the context of MISO wiretap channel with delayed CSIT and alternating topology, to the two-user MISO broadcast channel; and, so we outline it briefly. We begin the proof as follows.
\setlength{\arraycolsep}{0.02em}
\begin{align}
\label{b1-bc}
 nR_{1}&= H(W_1|W_2, z^n, \dv{S}^n ) \notag\\
&\overset{(a)}{\le} I(W_1; y^n | W_2,\dv{S}^n)- I(W_1; z^n| W_2,\dv{S}^n)+n\epsilon_{n} \notag\\
&\overset{(b)}{\le} h(z^n |W_2, \dv{S}^n)+n\lambda_{1 \alpha}(1-\alpha)\log(\rho)+n\epsilon_{n}
\end{align}
\noindent where $\epsilon_n \rightarrow 0$ as $n \rightarrow \infty$;
$(a)$ follows from Fano's inequality, $(b)$ follows by following similar steps leading to~\eqref{b1-wt} and by replacing $W$ with $W_1$.

We can also bound $R_1$ as follows.
\begin{align}
\label{b2-bc}
 nR_{1}&\le   I(W_1; y^n | \dv{S}^n)- I(W_1; z^n| \dv{S}^n)+n\epsilon_{n} \notag\\
&\overset{(c)}{\le} h(y^n | \dv{S}^n) - \frac{1}{2}h(z^n |\dv{S}^n)+\frac{n\lambda_{\alpha 1}(1-\alpha)}{2}\log(\rho)+n\epsilon_{n}
.\end{align}
\noindent where $(c)$ follows by similar steps leading to~\eqref{b2-wt} and  by replacing $W$ with $W_1$.

Then combining these two upper bounds in \eqref{b1-bc} and \eqref{b2-bc}, we get
\begin{align}
\label{c1-bc}
nR_{1}&\le \min\big\{h(z^n | W_2, \dv{S}^n)+n\lambda_{1\alpha}(1-\alpha)\log(\rho), h(y^n | \dv{S}^n) -\frac{1}{2}h(z^n |\dv{S}^n)+\frac{n\lambda_{\alpha 1}(1-\alpha)}{2}\log(\rho)\big\} +n\epsilon_{n}.
\end{align}
We can bound the rate $R_2$ as follows.
\begin{align}
\label{c2-bc}
nR_{2}&= H(W_2|\dv{S}^n)\notag\\
&= I(W_2;z^n|\dv{S}^n)+H(W_2|\dv{S}^n)\notag\\
&\overset{(d)}{\le} I(W_2;z^n|\dv{S}^n)+n\epsilon_{n}\notag\\
&= h(z^n | \dv{S}^n)-h(z^n |W_2, \dv{S}^n) +n\epsilon_{n}
\end{align}

\noindent where $(d)$ follows from Fano's inequality.

Next, by scaling~\eqref{c1-bc} with 3 and combining it with~\eqref{c2-bc}, we obtain
\begin{align}
& n(3R_1+R_{2})\notag\\&\le \min\{2h(z^n | W_2, \dv{S}^n)+h(z^n | \dv{S}^n)+3n\lambda_{1\alpha}(1-\alpha)\log(\rho), 3h(y^n | \dv{S}^n) -\frac{1}{2}h(z^n |\dv{S}^n)-h(z^n |W_2, \dv{S}^n)\notag\\& \:\:\:+\frac{3n\lambda_{\alpha 1}(1-\alpha)}{2}\log(\rho)\}+n{\epsilon}'_{n}\notag\\
\label{m1bc-mid}
&= \max_{\beta}\min\{\beta+3n\lambda_{1\alpha}(1-\alpha)\log(\rho), 3h(y^n | \dv{S}^n) -\frac{\beta}{2}+\frac{3n\lambda_{\alpha 1}(1-\alpha)}{2} \log(\rho)\} +n{\epsilon}'_{n}\\
&\overset{(e)}{\le} \max_{h(y^n)} 2h(y^n|\dv{S}^n)+n(1-\alpha)(\lambda_{1\alpha}+\lambda_{\alpha 1})\log(\rho)+n{\epsilon}'_{n}\notag\\\label{b2-1}
&\le\big[(3-\alpha)\lambda_{1 \alpha}+2(\lambda_{11}+\alpha\lambda_{\alpha\alpha})+(1+\alpha)\lambda_{\alpha 1}\big]n\log(\rho) +n{\epsilon}'_{n}
\end{align}
where we define $\beta:= 2h(z^n | W_2,\dv{S}^n)+h(z^n | \dv{S}^n)$; and $(e)$ follows by maximizing~\eqref{m1bc-mid} with respect to $\beta$.

Due to the symmetry of the problem, and following similar steps leading to~\eqref{b2-1}, we get 
\begin{align}
\label{m21-bc}
 n(R_1+3R_{2})&\le \max_{h(z^n)} 2h(z^n|\dv{S}^n)+n(1-\alpha)(\lambda_{1\alpha}+\lambda_{\alpha 1})\log(\rho)+n{\epsilon}'_{n} \\\label{m22-bc}
&\overset{(f)}{\le}\big[(3-\alpha)\lambda_{\alpha 1}+2(\lambda_{11}+\alpha\lambda_{\alpha\alpha})+(1+\alpha)\lambda_{1\alpha }\big]n\log(\rho) +n{\epsilon}'_{n}
\end{align}
where $(f)$ follows by maximizing~\eqref{m21-bc} with respect to $h(z^n)$ and due to the fact that conditioning reduces entropy. Then, dividing both sides by $n\log(\rho)$ and  taking $\lim \rho \rightarrow \infty$ and $\lim n \rightarrow \infty$, in~\eqref{b2-1} and~\eqref{m22-bc}, we get~\eqref{m1-bc} and~\eqref{m2-bc}, respectively.

This concludes the proof.

\section{Coding Scheme Achieving $\big(\frac{2}{3+\alpha},\frac{\alpha(1+\alpha)}{3+\alpha}\big)$ GSDoF Pair in Proposition~\ref{prop2} }
\label{app-4}
\iffalse
\begin{figure}
\psfragscanon
\centering
\psfrag{a}{$\rho$}
\psfrag{j}{$\rho^{1-\alpha}$}
\psfrag{i}{\hspace{-.5em}$\rho^{\alpha}$}
\psfrag{b}{$\tc{green}{v_1}$}
\psfrag{c}{\hspace{-.5em}$\tc{red}{\dv{h}_1\dv{u}}$}
\psfrag{d}{\hspace{-.25em}$\tc{red}{\dv{g}_1\dv{u}}$}
\psfrag{e}{\hspace{-.25em}$\dv{h}_2\tc{green}{\dv{v}}$}
\psfrag{+}{\hspace{.25em}$+$}
\psfrag{f}{\hspace{-1em}$h_{21}\tc{red}{\dv{h}_1\dv{u}}$}
\psfrag{g}{\hspace{-.25em}$\dv{g}_2\tc{green}{\dv{v}}$}
\psfrag{h}{\hspace{- 1em}$g_{21}\tc{red}{\dv{h}_1\dv{u}}$}
\psfrag{k}[c][c]{$\dv{h}_3\tc{brown}{\dv{w}}$}
\psfrag{l}[c][c]{\:\:\:$h_{31}\tc{red}{\dv{g}_1\dv{u}}$}
\psfrag{m}[c][c]{$\dv{g}_2\tc{brown}{\dv{w}}$}
\psfrag{n}{\hspace{- 1.1em}$g_{31}\tc{red}{\dv{g}_1\dv{u}}$}
\psfrag{o}[c][c]{$\tc{blue}{c}$}
\psfrag{p}[c][c]{$\tc{blue}{c}$}
\psfrag{q}[c][c]{\:\:$y_3'$}
\psfrag{s}{$\tc{green}{v_4}$}
\psfrag{t}{$\tc{green}{v_5}$}
\psfrag{t1}{$t_1=T$}
\psfrag{t2}{$t=2$}
\psfrag{t3}{$t=3$}
\psfrag{t4}{$t=4$}
\psfrag{R1}{\hspace{-2.5em}Received power levels at Rx$_1$}
\psfrag{R2}{\hspace{-2.5em}Received power levels at Rx$_2$}
\includegraphics[scale=1.15]{Power-levels-asym-sec-base.eps}
\caption{Received power levels at receiver 1 and receiver 2 with fixed topology $(\lambda_{1\alpha}:=1)$.}
\psfragscanoff
\label{power-level-prop-2-base}
\end{figure}
\fi
We now provide the proof of the coding scheme which gives the GSDoF pair  $(2/(3+\alpha),\alpha(1+\alpha)/(3+\alpha))$. The transmission scheme consists of four phases. \iffalse, each comprising of only one time slot.  In this scheme, transmitter wants to send five symbols  $(v_1,v_2,v_3,v_4,v_5)$ to receiver 1 and wishes to conceal them from receiver 2; and, two symbols $(w_1,w_2)$ to receiver 2 and wishes to conceal them from receiver 1, respectively. \fi 

\vspace{.5em}
\subsubsection{Phase 1}
In this phase communication takes place in $T_1$ channel uses, where the transmitter injects artificial noise from both antennas. Let $\dv{u}:=[\dv{u}_{1},\hdots,\dv{u}_{T_1}]^T$, where $\dv{u}_{t}= [u_{t,1}, u_{t,2}]^T$ $\forall\: t \in T_1$ denotes the noise injected by the transmitter. The channel input-output relationship is given by
 \begin{subequations}
 \begin{align}
 \label{n1}
\dv{y}_1 & =\underbrace{\sqrt{\rho}\tilde{{\dv{h}}}_1 {\dv{u}}}_{\mc O (\rho)}, \\
 \label{n2}
 \dv{z}_1 &= \underbrace{\sqrt{\rho^{\alpha}}\tilde{{\dv{g}}}_1{\dv{u}}}_{\mc O(\rho^{\alpha})}
 \end{align}
 \end{subequations}
where $\tilde{{\dv{h}}}_1 = \text{diag}(\{{{\tf{h}}}_t\}) \in \mb{C}^{T_1 \times 2T_1}$, $\tilde{{\dv{g}}}_1 = \text{diag}(\{{{\tf{g}}}_t\}) \in \mb{C}^{T_1 \times 2T_1}$, ${\dv{y}}_{1}  \in \mb{C}^{T_1}$ and ${\dv{z}}_{1} \in \mb{C}^{T_1}$,  for $t=1,\hdots,T_1$. At the end of phase 1, both receivers feed back the past CSI to the transmitter. By the help of past CSI, the transmitter can learn the channel outputs at both receivers.
\vspace{.5em}
\subsubsection{Phase 2}
In this phase communication takes place in $T_1$ channel uses. 
The transmitter sends fresh symbols  $\dv{v}:=[\dv{v}_{1},\hdots,\dv{v}_{T_1}]^T$, where  $\dv{v}_{t}= [v_{t,1}, v_{t,2}]^T$ $\forall\: t \in T_1$ to receiver 1 along with the past channel output at receiver 1 in phase 1 as
\begin{eqnarray}
 {\dv{x}}_2 = \dv{v}+\Theta_1\dv{y}_1
 \end{eqnarray}
 where $\Theta_1 \in  \mb{C}^{2T_1 \times T_1}$ is a matrix, that is assumed to be known at all nodes. The channel input-output relationship is given by
\begin{subequations}
 \begin{align}
 \label{r21}
\dv{y}_2 & =\underbrace{\sqrt{\rho}\tilde{{\dv{h}}}_2 (\dv{v}+\Theta_1 \dv{y}_1)}_{\mc O (\rho)}, \\
 \label{r22}
 \dv{z}_2 &= \underbrace{\sqrt{\rho^{\alpha}}\tilde{{\dv{g}}}_2(\dv{v}+\Theta_1\dv{y}_1)}_{\mc O(\rho^{\alpha})}
 \end{align}
 \end{subequations}
where $\tilde{{\dv{h}}}_2 = \text{diag}(\{{{\tf{h}}}_t\}) \in \mb{C}^{T_1 \times 2T_1}$, $\tilde{{\dv{g}}}_2 = \text{diag}(\{{{\tf{g}}}_t\}) \in \mb{C}^{T_1 \times 2T_1}$, ${\dv{y}}_{2}  \in \mb{C}^{T_1}$ and ${\dv{z}}_{2} \in \mb{C}^{T_1}$,  for $t=1,\hdots,T_1$. At the end of phase 2, both receivers feed back the past CSI to the transmitter. Since the receiver 1 knows the CSI $(\tilde{{\dv{h}}}_2)$ and the channel output at receiver 1 in phase 1 $(\dv{y}_1)$, it subtracts out the contribution of $\dv{y}_1$ from $\dv{y}_2$ to obtain $T_1$ equations with $2 T_1$ $\dv{v}$-variables and requires $T_1$ extra equations being available as side information at receiver 2 to decode the intended variables. Notice that the side information at receiver 2 is available at a reduced power level ($\mc O(\rho^{\alpha})$) compared to the receiver 1. 

\subsubsection{Phase 3}
This phase is similar to phase 2, with the roles of receiver 1 and receiver 2 being reversed.
In this phase communication takes place in $T_2 :=\alpha T_1$ channel uses. The transmitter sends confidential symbols $\dv{w}:=[\dv{w}_{1},\hdots,\dv{w}_{T_2}]^T$, where  $\dv{w}_{t}= [w_{t,1}, w_{t,2}]^T$ $\forall\: t \in T_2$ to receiver 2 along with the past channel output at receiver 2 as
\begin{eqnarray}
 {\dv{x}}_3 = \dv{w}+\Theta_2 {\dv{z}}_1
 \end{eqnarray}
where  $\Theta_2 \in  \mb{C}^{2T_2 \times T_1}$ is a matrix, that is assumed to be known at all nodes. The channel input-output relationship is given by
\begin{subequations}
 \begin{align}
 \label{r31}
\dv{y}_3 & =\underbrace{\sqrt{\rho}\tilde{{\dv{h}}}_3 (\dv{w}+\Theta_2{\dv{z}}_1)}_{\mc O (\rho)}, \\
 \label{r32}
 \dv{z}_3 &= \underbrace{\sqrt{\rho^{\alpha}}\tilde{{\dv{g}}}_3(\dv{w}+\Theta_2{\dv{z}}_1)}_{\mc O(\rho^{\alpha})}
 \end{align}
 \end{subequations}
where $\tilde{{\dv{h}}}_3 = \text{diag}(\{{{\tf{h}}}_t\}) \in \mb{C}^{T_2 \times 2T_2}$, $\tilde{{\dv{g}}}_3 = \text{diag}(\{{{\tf{g}}}_t\}) \in \mb{C}^{T_2 \times 2T_2}$, ${\dv{y}}_{3}  \in \mb{C}^{T_2}$ and ${\dv{z}}_{3} \in \mb{C}^{T_2}$,  for $t=1,\hdots,T_2$. At the end of phase 3, both receivers feed back the past CSI to the transmitter. Since the receiver 2 knows the CSI $(\tilde{{\dv{g}}}_3)$ and the channel output at receiver 1 in phase 1 $(\dv{z}_1)$, it subtracts out the contribution of ${\dv{z}}_1$ from $\dv{z}_3$ to obtain $T_2$ equations with $2 T_2$ $\dv{w}$-variables and requires $T_2$ equations being available as side information at receiver 1 --- however at a higher  power level ($\mc O(\rho)$) compared to the receiver 2. 
\vspace{.5em}
\subsubsection{Phase 4}
\label{subsection-phase4}
At the end of phase 3, receiver 1 requires side information $(\dv z_2)$  at receiver 2 in phase 2 and receiver 2 requires side information $(\dv y_3)$ at receiver 1 in phase 3 to successfully decode the intended variables. The transmitter can easily \ti{learn} them by means of past CSI; and, the goal of this phase is to communicate these side informations for interference alignment \`{a}-la MAT scheme~\cite{M-AT12}. The key ingredients of the transmission scheme are, 1) by opposition to classical MAT scheme where side information  is conveyed in an analog manner, digitized side information is multicasted, and 2) transmission of fresh information to receiver 1. Recall that, at the end of phase 3, the side information at receiver 2 in phase 2 is available at a reduced power level $\mc O (\rho^\alpha)$ compared to the side information at receiver 1 $\mc O (\rho)$ in phase 3. After learning the side information, the transmitter perform following operations --- it quantizes the channel output at receiver 2 $(\dv{z}_2)$ into  $\alpha T_1 \log(\rho) + o(\log \rho)$ bits  within bounded noise distortion. A similar operation is performed at the receiver 1 channel output in phase 3 --- the transmitter first generates  $(\dv{y}_3':=\Psi_1\dv{y}_3)$ where $\Psi_1\in \mb{C}^{T_1\times T_2}$ and is assumed to be known everywhere, and quantize it to $ T_2 \log(\rho) + o(\log \rho)$ bits  within bounded noise distortion. Since $T_2:=\alpha T_1$, the transmitter performs a bit wise XOR operation to generate  $\alpha T_1 \log(\rho) + o(\log \rho)$ bits which are then mapped to a common message $\{c_t\}$ where $c_t \in \mc C = \{1,\hdots,\rho^{\alpha}\}$ $\forall$ $t \in T_1$ and transmits it with fresh information ($v_{t,3}$) for receiver 1 as 
\begin{eqnarray}
{\dv{x}}_{t,4} = \left [
\begin{matrix}
c_t+v_{t,3}\rho^{-\alpha/2}\\
\phi
\end{matrix} \right ]
\end{eqnarray}
where $\mb{E}[||v_{t,3}||^2] \doteq 1$ and $\mb{E}[||c_t||^2] \doteq 1$, $\forall$ $t \in T_1$.  At the end of phase 4, the channel input-output relationship is given by
\begin{subequations}
\begin{align}
\dv y_4 &= \underbrace{\sqrt{\rho} \dv {h_{41}} \dv{c}}_{\mc O (\rho )}+\underbrace{\sqrt{\rho^{(1-\alpha)}} \dv{h_{41}} \dv{v_3}}_{\mc O (\rho^{1-\alpha} )}\\
\dv z_4 &= \underbrace{\sqrt{\rho^{\alpha}} \dv {g_{41}} \dv{c}}_{\mc O (\rho )}+\underbrace{\sqrt{\rho^{0}} \dv{g_{41}} \dv{v_3}}_{\mc O (\rho^{0} )}
\end{align}
\end{subequations}
where ${{\dv{h}}}_{41} = \text{diag}(\{{{\tf{h}}}_t\}) \in \mb{C}^{T_1 \times T_1}$, ${{\dv{g}}}_{41} = \text{diag}(\{{{\tf{g}}}_t\}) \in \mb{C}^{T_1 \times  T_1}$, ${\dv{c}} = [c_1,\hdots,c_{T_1}]^T$, ${\dv{v}}_3 = [v_{1,3},\hdots,v_{T_1,3}]^T$,   ${\dv{y}}_{4}  \in \mb{C}^{T_1}$ and ${\dv{z}}_{4} \in \mb{C}^{T_1}$,  for $t=1,\hdots,T_1$.
At the end of phase 4, receiver 2 gets the confidential symbols $\dv{v}_3$ intended for receiver 1 at noise floor level and is unable to decode it.
Receiver 1 first  constructs $\dv h_{41}\dv c$ from the channel output $\dv y_4$ by treating $\dv v_3$ as noise, and, afterwards can easily decode $\dv v_3$. Through straightforward algebra, it can be readily shown that $(1-\alpha)T_1\log(\rho)$ bits are securely conveyed by symbol $\dv v_3$ to receiver 1. By using $\dv c$, both receivers can learn the side information required to decode the intended symbols as follows. After decoding $\dv c$ receiver 1 performs following operations --- 1) since the receiver 1 knows $\dv{y}_3$ and  $\Psi_1$, it first performs XOR operation between the quantized version of $\dv y_3'$  and $\dv c$ to recover the side information ($\dv z_2$) within bounded noise distortion, and then 2) it subtracts out the contribution of $\dv y_1$ from $(\dv y_2,\dv z_2)$ to decode $\dv{v}$ through channel inversion.  Due to the symmetry of the problem receiver 2 can also perform similar operations to first recover side information $\dv y_3$ and from $(\dv z_3, \dv y_3)$ decodes $\dv{w}$ through channel inversion.

\vspace{.5em}
\subsubsection{Equivocation Analysis}
\label{equiv}
 We can write the channel input-output relationship as
\begin{align}
{\dv{y}}&:=  \underbrace{\left [
\begin{matrix}
\sqrt{\rho}  \tilde{\dv{h}}_2  & \sqrt{\rho} \tilde{\dv{h}}_2\Theta_1& \dv 0 \:\:&\:\: \dv 0 \\
\sqrt{\rho^\alpha } {\dv{h}}_{41}\tilde{\dv{g}}_2 &\:\: \sqrt{\rho^\alpha}  {\dv{h}}_{41}\tilde{\dv{g}}_2\Theta_1 &\:\:\sqrt{\rho }  {\dv{h}}_{41}\Psi_1 &\:\: \sqrt{\rho^{1-\alpha} }\tilde {\dv{h}}_{41}\\
\dv 0 & \sqrt{\rho} \dv {I}_{T_1} & \dv 0 & \dv 0\\
\dv 0 & \dv 0 & \sqrt{\rho} \dv{I}_{T_2} & \dv 0\\
\end{matrix} \right ]}_{\dv H \in \mb{C}^{3T_1+T_2 \times 4T_1+T_2}}\left [
\begin{matrix}
\dv{v}\\
\tilde{\dv{h}}_{1}{\dv{u}}  \\
\tilde{{\dv{h}}}_3 {\dv{w}}+ \tilde{\dv{h}}_{3}\Theta_2{\tilde{\dv{g}}}_1 {\dv{u}} \\
\dv{v}_3
\end{matrix} \right ],\\
{\dv{z}}&:=  \underbrace{\left [
\begin{matrix}
\dv 0  & \sqrt{\rho^{\alpha}} \dv{I}_{T_1}& \dv 0 & \dv{0} \\
\dv 0  & 0 &\sqrt{\rho^{\alpha}}\dv{I}_{T_1} & \dv{0}\\
\sqrt{\rho^{\alpha}}\dv{g}_3 & \sqrt{\rho^{\alpha}}\dv{g}_{3}\Theta_2  & \dv {0} & \dv{0}\\
   \sqrt{\rho} \dv {g}_{41}\Psi_1\tilde{\dv{h}}_3 & \:\: \sqrt{\rho} \dv {g}_{41}\Psi_1\tilde{\dv{h}}_3\Theta_2 &\:\:  \sqrt{\rho^{\alpha}} \dv g_{41}&\:\:\sqrt{\rho^{0}}\dv{g}_{41} \\
\end{matrix} \right ]}_{\dv G \in \mb{C}^{3T_1+T_2 \times 3T_1+2T_2}}\left [
\begin{matrix}
\dv{w}\\
\tilde{\dv{g}}_{1}{\dv{u}}  \\
\tilde{{\dv{g}}}_2 {\dv{v}}+ \tilde{{\dv{g}}}_{2}\Theta_1{{\dv{h}}}_1 {\dv{u}} \\
\dv{v}_3
\end{matrix} \right ].
\end{align} 

The information rate to receiver 1 is bounded by 
\begin{align}
I(\dv{v}, \dv{v}_3;\dv{y}|\dv{S}^n) &= I({{\dv{v}}},\dv{v}_3;\dv{y}_1,\dv {y}_3|\dv{S}^n)+ I({{\dv{v}}};\dv {y}_2, \dv {y}_4|{{\dv{v}}}_3,\dv {y}_1, \dv {y}_3, \dv{S}^n)+I({{\dv{v}}}_3;\dv {y}_2, \dv {y}_4|\dv {y}_1, \dv {y}_3, \dv{S}^n)\notag\\
& \overset{(a)}{=} I({{\dv{v}}};\dv {y}_2,\dv{y}_4|{{\dv{v}}}_3,\dv {y}_1,\dv {y}_3,\dv{S}^n)+(1-\alpha)T_1 \log (\rho) \notag\\
& = (1+\alpha)T_1\log(\rho)+(1-\alpha)T_1 \log (\rho) 
\end{align}
where $(a)$ follows  due to the independence of  $(\dv{v},\dv{v}_3)$ and $(y_1,y_3)$.

We can bound the information leakage to receiver 2 as
\begin{align}
I(\dv{v}, \dv{v}_3;\dv{z}|\dv{w},\dv{S}^n) &= I(\dv{v};\dv{z}|\dv{v}_3, \dv{w},\dv{S}^n)+\underbrace {I(\dv{v}_3;\dv{z}|\dv{w},\dv{S}^n)}_{o (\log (\rho))}\notag\\
&\le I(\tilde{\dv{g}}_2\dv{v}, \dv{u};\dv{z}|\dv{v}_3, \dv{w},\dv{S}^n)-I(\dv{u};\dv{z}|\tilde{\dv{g}}_2\dv{v},\dv{v}_3, \dv{w},\dv{S}^n)+o (\log (\rho))\notag\\
&\le I(\tilde{{\dv{g}}}_2 {\dv{v}}+ \tilde{{\dv{g}}}_{2}\Theta_1{{\dv{h}}}_1 {\dv{u}}, \dv{u};\dv{z}|\dv{v}_3, \dv{w},\dv{S}^n)- I(\dv{u};\dv{z}|\tilde{\dv{g}}_2\dv{v},\dv{v}_3, \dv{w},\dv{S}^n)+ o (\log (\rho)) \notag\\
& \overset{(b)}{=}2\alpha T_1\log(\rho)-2\alpha T_1\log(\rho)+o(\log (\rho)) \notag\\
& = o(\log (\rho))
\end{align}
where $(b)$ follows from \cite[Lemma 2]{YKPS11}.

\noindent From the analysis above, it can be readily seen that $(1+\alpha)T_1\log (\rho)$ bits are securely send by $\dv{v}$; and $(1-\alpha)T_1\log (\rho) $ bits are securely send by $\dv{v}_3$, to receiver 1 over a total of $3T_1+T_2=(3+\alpha)T_1$ time slots, yielding $d_1=\frac{2}{3+\alpha}$ GSDoF at receiver 1. Similar analysis shows  that $(1+\alpha)T_2\log{(\rho)}$ bits are  securely transmitted via $\dv{w}$ to receiver 2 over a total of $(3+\alpha)T_1$ time slots, yielding $d_{2}=\frac{\alpha(1+\alpha)}{3+\alpha}$ GSDoF at receiver 2.

This concludes the proof.

\section{Coding Scheme Achieving $(\frac{1+\alpha}{4},\frac{1}{2})$ SDoF Pair in Proposition~\ref{prop4} }
\label{app-6}
In this coding scheme communication takes place in four phases, each consisting of only one time slot. The transmitter alternate between different states and uses $\lambda_{1 \alpha}$ state at $t=1, 2,$ and $\lambda_{\alpha 1}$ state at $t=3, 4,$ respectively. In this scheme, transmitter wants to send two symbols $(v_1,v_2)$ to receiver 1 that are meant to be kept secret from  receiver 2 and three symbols $(w_1,w_2,w_3)$ to receiver 2  that are meant to be kept secret from receiver 1.

In the first and second phase, topology state  $\lambda_{1 \alpha}$  occurs.  The transmission scheme in this case is similar to that in phase 1 and 2 of the coding scheme of Appendix~\ref{app-4}. The channel inputs-outputs relationship at receiver 1 $(y_1,y_2)$ and receiver 2 $(z_1,z_2)$ are given by~\eqref{n1}, \eqref{r21} and~\eqref{n2}, \eqref{r22}, respectively, where $T_1:=1$ and $\Theta_1 := [1, 0]^T$. Figure~\ref{Power-levels-sym-secrecy} illustrates the power levels at both receivers. \iffalse At the end of this phase, receiver 1 wants to decode  the confidential symbols $(v_2,v_3)$ sent by the transmitter and requires one extra equation to decode them. The receiver 2 gets a linear combination of fresh information intended for receiver 1 embedded in with artificial noise. If this equation $(z_2)$ can be conveyed to receiver 1, it suffices to decode symbols $(\dv {v}=[v_2,v_3]^T)$. This equation is conveyed by transmitter is phase 4.\fi
\begin{figure}
\psfragscanon
\centering
\psfrag{a}{$\rho$}
\psfrag{j}{$\rho^{1-\alpha}$}
\psfrag{r}{$\rho^{0}$}
\psfrag{i}{\hspace{-.5em}$\rho^{\alpha}$}
\psfrag{b}{$\tc{green}{v_1}$}
\psfrag{c}{\hspace{-.5em}$\tc{red}{\dv{h}_1\dv{u}}$}
\psfrag{d}{\hspace{-.25em}$\tc{red}{\dv{g}_1\dv{u}}$}
\psfrag{e}{\hspace{-.25em}$\dv{h}_2\tc{green}{\dv{v}}$}
\psfrag{+}{\hspace{.25em}$+$}
\psfrag{f}{\hspace{-.75em}$h_{21}\tc{red}{\dv{h}_1\dv{u}}$}
\psfrag{g}{\hspace{-.25em}$\dv{g}_2\tc{green}{\dv{v}}$}
\psfrag{h}{\hspace{- 1.2em}$g_{21}\tc{red}{\dv{h}_1\dv{u}}$}
\psfrag{m}[c][c]{$\dv{h}_3\tc{brown}{\dv{w}}$}
\psfrag{n}[c][c]{\:\:\:$h_{31}\tc{red}{\dv{g}_1\dv{u}}$}
\psfrag{k}[c][c]{$\dv{g}_3\tc{brown}{\dv{w}}$}
\psfrag{l}{\hspace{- 1.1em}$g_{31}\tc{red}{\dv{g}_1\dv{u}}$}
\psfrag{o}[c][c]{$\tc{blue}{c}$}
\psfrag{p}[c][c]{$\tc{blue}{c}$}
\psfrag{t}{$\tc{brown}{w_3}$}
\psfrag{t1}{$t=1$}
\psfrag{t2}{$t=2$}
\psfrag{t3}{$t=3$}
\psfrag{t4}{$t=4$}
\psfrag{R1}{\hspace{-2.5em}Received power levels at Rx$_1$}
\psfrag{R2}{\hspace{-2.5em}Received power levels at Rx$_2$}
\includegraphics[scale=1]{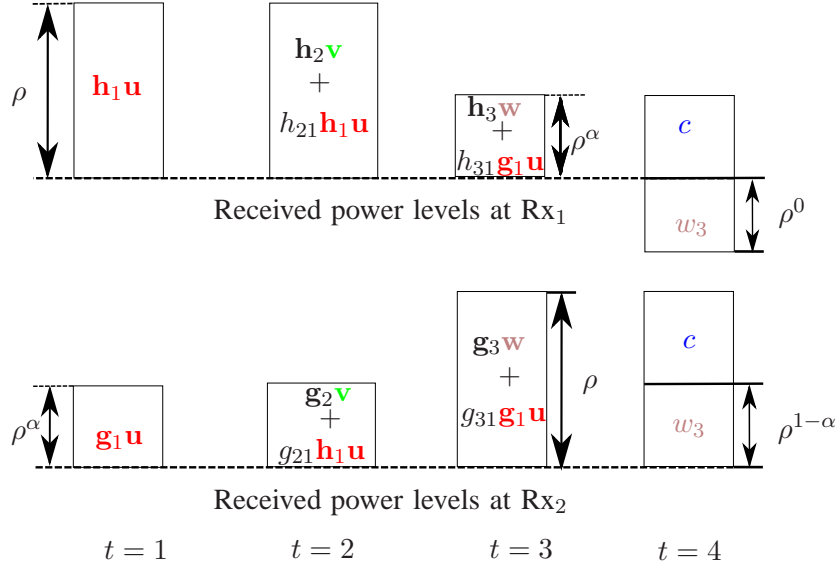}
\caption{Received power levels at receiver 1 and receiver 2 with alternating topology $(\lambda_{1\alpha},\lambda_{\alpha 1}):=(\frac{1}{2},\frac{1}{2})$.}
\psfragscanoff
\label{Power-levels-sym-secrecy}
\end{figure}
In the third phase, the transmitter switches to $\lambda_{\alpha 1}$ topology state. Recall that, in this state receiver 2 is comparatively stronger than receiver 1. The transmission scheme in this phase follows by reversing the roles of receiver 1 and 2, respectively. In this phase, transmitter sends fresh information ($\dv{w}:=[w_1, w_2]^T$) to receiver 2 along with a linear combination of channel output $({{\dv{g}}}_1 {\dv{u}})$ at receiver 2 during the first phase, i.e.,
\begin{eqnarray}
{\dv{x}}_3 = \left [
\begin{matrix}
w_{1}\\
w_{2}
\end{matrix} \right ] + \left [
\begin{matrix}
{{\dv{g}}}_1 {\dv{u}} \\
\phi
\end{matrix} \right ]. \:\:\:\:\:\:
\end{eqnarray}
The channel input-output  relationship is given by
\begin{subequations}
\begin{align}
 y_3 &= \sqrt{\rho^{\alpha}}({{\dv{h}}}_3 {\dv{w}}+ h_{31}{{\dv{g}}}_1 {\dv{u}})\\
z_3 &=\sqrt{\rho}({{\dv{g}}}_3 {\dv{w}}+ g_{31}{{\dv{g}}}_1 {\dv{u}}).
\end{align}
\end{subequations}
At the end of phase 3, receiver 2 gets the confidential symbols ($\dv{w}:=[w_1, w_2]^T$) embedded in with artificial noise. Since receiver 2 knows the CSI $({{\dv{g}}}_3)$, and also the channel output $z_1$ from phase 1, it subtracts out the contribution of $z_1$ from the channel output $z_3$, to obtain one equation with two unknowns (${\dv{w}}:=[w_1,w_2]^T$) and requires one extra equation to successfully decode the intended variables being available as side information at receiver 1. This equation is conveyed by the transmitter to receiver 2 in phase 4.

In phase 4, the channel remains in $\lambda_{\alpha 1}$ topology state. At the end of phase 3, receiver 1 requires side information $z_2$  at receiver 2 in phase 2 and receiver 2 requires side information $y_3$ at receiver 1 in phase 3 to successfully decode the intended variables. Notice that the side information available at the unintended receivers are available at reduced power levels $(\mc O ( \rho^{\alpha}))$. In this phase, after learning $(y_3,z_2)$, transmitter first generates a new symbol $s:=y_3+z_2$ where $\mb{E}[||s||^2]=\mc O(\rho^\alpha)$. Afterwards, it quantizes $
s=\hat{s}+\Delta$, 
where $\Delta$ is the quantization error, $\hat{s}$ is the quantized value and contains
\begin{eqnarray}
R_{\hat s} &=& I(s;\hat{s})\notag\\
&\overset{(a)}{\approx}& \alpha \log(\rho)\:\: \text{bits}
\end{eqnarray}
where $(a)$ follows because the quantization error $\mathbb{E}[\Delta^2]=\mathbb{E}[||\hat{s}-s||^2]$ is bounded and does not scale asymptotically with $\log(\rho)$. 
The transmitter then maps the quantization index $\hat{s}$ to a common symbol $c$ using a Gaussian codebook, where $c \in \mc {C} = \{1,\hdots,\rho^{\alpha}\}$, and transmits it with fresh information ($w_3$) for receiver 2 as 
\begin{eqnarray}
{\dv{x}}_4 = \left [
\begin{matrix}
c+w_3\rho^{-\alpha/2}\\
\phi
\end{matrix} \right ]
\end{eqnarray}
where $\mb{E}[||w_3||^2] \doteq 1$ and $\mb{E}[||c||^2] \doteq 1$.
 At the end of phase 4, the channel input-output relationship is given by
\begin{subequations}
\begin{align}
y_4 &= \underbrace{\sqrt{\rho^{\alpha}} h_{41} c}_{\mc O (\rho^\alpha )}+\underbrace{\sqrt{\rho^{0}} h_{41} w_3}_{\mc O (\rho^{0} )}\\
z_4 &=  \underbrace{\sqrt{\rho}  g_{41} c}_{\mc O (\rho)} +\underbrace{\sqrt{\rho^{(1-\alpha)}} g_{41} w_3}_{\mc O (\rho^{1-\alpha} )}.
\end{align}
\end{subequations}
At the end of phase 4, receiver 1 gets the confidential symbol ${w}_3$ intended for receiver 2 at noise floor level and is unable to decode it.
Receiver 2 can easily reconstruct $c$ and subtracts out its contribution from $z_4$ to decode the confidential symbol $w_3$ through channel inversion. Subsequently, by using $c$ and following steps as mentioned before in subsection~\ref{subsection-phase4}, receiver 1 can learn the side information $z_2$ and receiver 2 can learn the side information $y_3$. Thus, with the help of $(y_1,y_2,z_2)$ receiver 1 can successfully decodes the symbols  $(v_1,v_2)$ that are intended to it. By using the side information $y_3$ and the channel outputs $(z_1,z_3)$, receiver 2    decodes the intended variables $(w_1,w_2)$.
 
Following steps similar to in Appendix~\ref{equiv}, it can be readily seen that $(1+\alpha)\log(\rho)$ bits are securely transmitted to receiver 1 via $\dv{v}$ over a total of $4$ time slots, yielding $d_1=\frac{1+\alpha}{4}$ SDoF at receiver 1. Due to the symmetry of the problem, it can be readily shown that $(1+\alpha+1-\alpha)\log(\rho)$ bits are transmitted securely to receiver 2 over a total of $4$ time slots, yielding $d_{2}=\frac{1}{2}$ SDoF at receiver 2.

This concludes the proof.

\section{Coding Scheme Achieving $(\frac{1}{2},\frac{1}{2})$ SDoF Pair in Proposition~\ref{prop5} }
The transmission scheme in this case is similar to the one in Proposition~\ref{prop2} and Proposition~\ref{prop1-new}, so we outline it briefly. The communication takes place in four phases, each consisting of only one time slot. In this scheme the transmitter alternate between different states and uses $\lambda_{1 \alpha}$ state at $t=1, 2,$ and $\lambda_{\alpha 1}$ state at $t=3, 4,$ respectively. The transmitter wants to send three symbols $(v_1,v_2,v_3)$ to receiver 1 that are meant to be kept secret from  receiver 1 and three symbols $(w_1,w_2,w_3)$ to receiver 2  that are meant to be kept secret from receiver 1. In the first phase, by utilizing the leverage due to the topology of the network, transmitter injects structured noise (see Proposition~\ref{prop1-new} for details) and a confidential symbol $v_1$ to receiver 1. Figure~\ref{Power-levels-sym-secrecy} illustrates the power levels at both receivers. The untended receiver obtains $v_1$ below noise floor level and is thus unable to decode it. The rest of the steps in the coding scheme are similar to in Appendix~\ref{app-4} and is omitted. Thus, at the end of four timeslot $(1-\alpha)\log(\rho)$ bits are securely send by $v_1$ and  $(1+\alpha)\log(\rho)$ bits are securely conveyed by  $\dv{v}:=[v_2,v_3]^T$ over a total of four time slots yielding $d_1=\frac{1}{2}$. Due to the symmetry of the problem the receiver 2 yields a GSDoF of $d_2=\frac{1}{2}$. 

This concludes the proof.

\begin{figure}
\psfragscanon
\centering
\psfrag{a}{$\rho$}
\psfrag{r}{$\rho^{0}$}
\psfrag{j}{$\rho^{1-\alpha}$}
\psfrag{i}{\hspace{-.5em}$\rho^{\alpha}$}
\psfrag{c}{$\tc{green}{v_1}$}
\psfrag{b}{\hspace{-.5em}$\tc{red}{\dv{h}_1\dv{u}}$}
\psfrag{d}{\hspace{-.25em}$\tc{red}{\dv{g}_1\dv{u}}$}
\psfrag{e}{\hspace{-.25em}$\dv{h}_2\tc{green}{\dv{v}}$}
\psfrag{+}{\hspace{.25em}$+$}
\psfrag{f}{\hspace{-.85em}$h_{21}\tc{red}{\dv{h}_1\dv{u}}$}
\psfrag{g}{\hspace{-.25em}$\dv{g}_2\tc{green}{\dv{v}}$}
\psfrag{h}{\hspace{- 1.2em}$g_{21}\tc{red}{\dv{h}_1\dv{u}}$}
\psfrag{m}[c][c]{$\dv{h}_3\tc{brown}{\dv{w}}$}
\psfrag{n}[c][c]{\:\:\:$h_{31}\tc{red}{\dv{g}_1\dv{u}}$}
\psfrag{k}[c][c]{$\dv{g}_3\tc{brown}{\dv{w}}$}
\psfrag{l}{\hspace{- 1.1em}$g_{31}\tc{red}{\dv{g}_1\dv{u}}$}
\psfrag{o}[c][c]{$\tc{blue}{c}$}
\psfrag{p}[c][c]{$\tc{blue}{c}$}
\psfrag{t}{$\tc{brown}{w_3}$}
\psfrag{t1}{$t=1$}
\psfrag{t2}{$t=2$}
\psfrag{t3}{$t=3$}
\psfrag{t4}{$t=4$}
\psfrag{R1}{\hspace{-2.5em}Received power levels at Rx$_1$}
\psfrag{R2}{\hspace{-2.5em}Received power levels at Rx$_2$}
\includegraphics[scale=1]{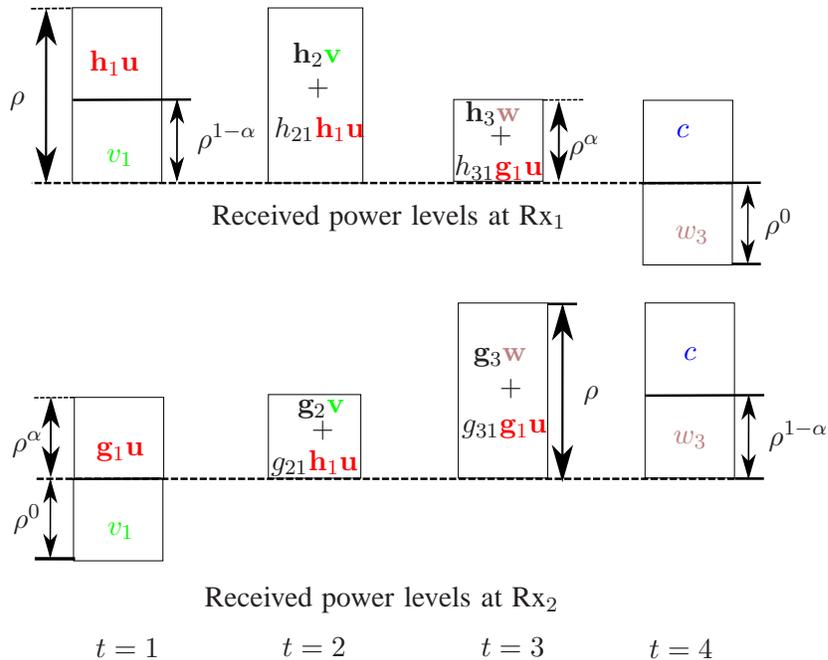}
\caption{Received power levels at receiver 1 and receiver 2 with alternating topology $(\lambda_{1\alpha},\lambda_{\alpha 1}):=(\frac{1}{2},\frac{1}{2})$.}
\psfragscanoff
\label{Power-levels-sym-secrecy}
\end{figure}
 
\section{Coding Scheme Achieving $(1-\frac{\alpha}{3},\frac{2\alpha}{3})$ DoF Pair in Theorem~\ref{prop3}}
\label{app-5}
In this scheme transmitter wants to send five symbols $(v_1,v_2,v_3,v_4,v_5)$ to receiver 1 and two symbols $(w_1,w_2)$ to receiver 2, respectively. 
\begin{figure}
\psfragscanon
\centering
\psfrag{a}{$\rho$}
\psfrag{j}{\hspace{-.25em}$\rho^{1-\alpha}$}
\psfrag{i}{\hspace{-.5em}$\rho^{\alpha}$}
\psfrag{n}{$\rho^{\alpha}$}
\psfrag{r}{\hspace{-.5em}$\rho^{0}$}
\psfrag{f}{\hspace{-.15em}$\tc{green} {v_3}$}
\psfrag{e}{$\tc{green}{v_4}$}
\psfrag{k}{$\tc{green}{v_5}$}
\psfrag{c}{\hspace{-.5em}$\dv{h}_1\tc{green}{\dv{v}}$}
\psfrag{g}{\hspace{-.25em}$\dv{g}_1\tc{green}{\dv{v}}$}
\psfrag{b}{\hspace{-.25em}$\dv{h}_2\tc{brown}{\dv{w}}$}
\psfrag{d}{\hspace{-.25em}$\dv{g}_2\tc{brown}{\dv{w}}$}
\psfrag{h}{\hspace{- 1.2em}$g_{21}\tc{red}{\dv{h}_1\dv{u}}$}
\psfrag{s}[c][c]{$ \tc{blue}{c}$}
\psfrag{t1}{$t=1$}
\psfrag{t2}{$t=2$}
\psfrag{t3}{$t=3$}
\psfrag{R1}{\hspace{-5em}Received power levels at Rx$_1$}
\psfrag{R2}{\hspace{-5em}Received power levels at Rx$_2$}
\includegraphics[scale=1]{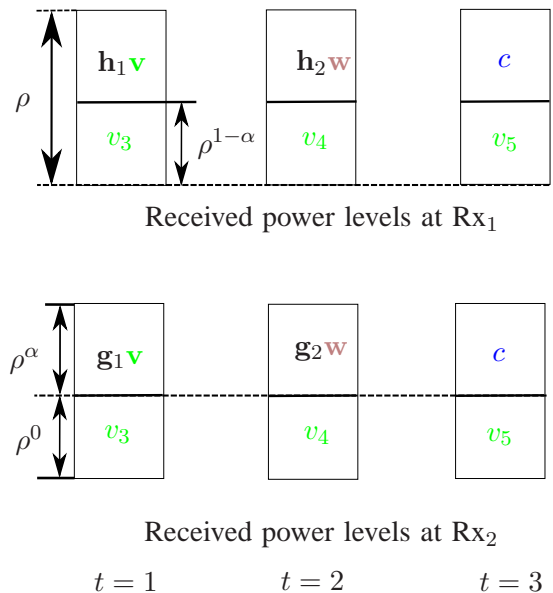}
\caption{Received power levels at receiver 1 and receiver 2.}
\psfragscanoff
\label{Power-levels-asym-without-secrecy}
\end{figure}
The transmission scheme consists of four timeslot. In the first time slot  the transmitter sends three symbols to receiver 1 as follows. By utilizing the topology of the network, the transmitter sends two symbols $\dv{v}:=[v_1, v_2]^T$ at the power level of $(\mc{O}(\rho^\alpha))$ and uses the remaining power to send a new symbol $v_3$ to receiver 1. As said before in Proposition~\ref{prop1-new}, the two symbols $\dv{v}$ are chosen from a lattice codebook while symbol $v_3$ is selected from a Gaussian codebook. More specifically, the transmitter sends
\begin{eqnarray}
{\dv{x}}_1 = \left [
\begin{matrix}
v_{1}\\
v_{2}
\end{matrix} \right ] + \left [
\begin{matrix}
v_3\rho^{-\alpha/2} \\
\phi
\end{matrix} \right ]. \:\:\:\:\:\:
\end{eqnarray}
The channel input-output relationship in terms of power levels is illustrated in Figure~\ref{Power-levels-asym-without-secrecy}. In the second timeslot, the transmitter sends fresh symbols $\dv{w}:=[w_1, w_2]^T$ to receiver 2 chosen from a lattice codebook along with a new symbol $v_4$ to receiver 1 --- selected from a Gaussian codebook, as follows
\begin{eqnarray}
{\dv{x}}_2 = \left [
\begin{matrix}
w_{1}\\
w_{2}
\end{matrix} \right ] + \left [
\begin{matrix}
v_4\rho^{-\alpha/2} \\
\phi
\end{matrix} \right ]. \:\:\:\:\:\:
\end{eqnarray}
At the end of the second time slot, the receiver 1 can easily construct $\dv{h}_1 \dv{v}$ and $\dv{h}_2 \dv{w}$, respectively; and, subsequently decodes $v_3$ and $v_4$, where each symbol contains $(1-\alpha)\log{(\rho)}$ bits. At the end of second timeslot, receiver 1 requires the channel output at receiver 1 in time slot 1 and receiver 2 requires the part of channel output at receiver 1 in timeslot 2 to decode the intended symbols. Due to the availability of delayed CSI, the transmitter can learn the side informations $(\dv{g}_1 \dv{v},\dv{h}_2 \dv{w})$ and in the third timeslot sends a linear combination of them along with a fresh symbol for receiver 1. Note that, since linear combination of  $\dv{g}_1 \dv{v}$ and $\dv{h}_2 \dv{w}$ is also a lattice point, so the transmitter first constructs $c:= \dv{g}_1 \dv{v}+\dv{h}_2 \dv{w}$ and sends
\begin{eqnarray}
{\dv{x}}_3 = \left [
\begin{matrix}
c\\
\phi
\end{matrix} \right ] + \left [
\begin{matrix}
v_5\rho^{-\alpha/2} \\
\phi
\end{matrix} \right ]. \:\:\:\:\:\:
\end{eqnarray}
At the end of third timeslot, receiver 1 can first compute $c$ and then by subtracting the contribution of $c$ from $y_3$ decodes $v_5$. Afterwards, from $c$ it recovers $\dv{g}_1\dv{v}$, and with the help of $(\dv{g}_1\dv{v},\dv{h}_1\dv{v})$ receiver 1 decodes $\dv{v}$ through channel inversion. Due to the symmetry of problem receiver 2 can also perform similar operations to decode $\dv{w}$.
At the end of transmission, $(2\alpha)\log(\rho)$ bits are send by $\dv {v}:=[v_1,v_2]$ and $(1-\alpha)\log(\rho)$ bits are send by each symbol $v_3$, $v_4$ and $v_5$ to receiver 1, respectively,  over a total of three time slots yielding a DoF of $1-\frac{\alpha}{3}$ at receiver 1. Similarly, $(2\alpha)\log(\rho)$ bits are send to receiver 2 via $\dv {w}:=[w_1,w_2]$ over a total of 3 timeslots yielding a DoF of $2\alpha/3$ at receiver 2.

This concludes the proof.

\bibliographystyle{IEEEtran} 
\bibliography{IEEEabrv,MISOBC-bib}
%\balance
\end{document}